\documentclass[aps]{revtex4}
\usepackage{amsmath,amssymb}
\usepackage{color,graphicx}
\usepackage{multirow}
\usepackage{amsmath,amssymb}
\usepackage{slashbox}
\newcommand{\be}{\begin{equation}}
\newcommand{\ee}{\end{equation}}
\newcommand{\calka}{\int\limits_0^\infty dr \int\limits_0^{2\pi} d\phi \int\limits_{0}^{\pi} d\theta}
\newcommand{\iksy}{x_1^2+x_2^2+x_3^2}

\begin{document}
\title{Ultrarelativistic bound states in  the shallow spherical well}
\author{Mariusz \.{Z}aba  and  Piotr Garbaczewski}
 \affiliation{Institute of Physics, University of Opole, 45-052 Opole, Poland}
\begin{abstract}
We determine approximate eigenvalues and eigenfunctions shapes for bound states in the $3D$
  shallow spherical ultrarelativistic well. Existence thresholds for the ground state and first
   excited states are identified, both in the purely radial and orbitally nontrivial cases.
 This contributes to an  understanding of how energy may be stored or accumulated  in the form of
 bound states of Schr\"{o}dinger - type  quantum systems that are devoid of any mass.
    \end{abstract}
\maketitle

\section{Motivation}

In the present paper the term ultrarelativistic refers to  the energy operator $\hat{H}_0 =  \hbar c \sqrt{- \Delta }$  in three space dimensions.
It is often regarded as the
  zero mass relative  of the quasirelativistic operator  $\hat{H}_m = \sqrt{-\hbar ^2c^2 \Delta + m^2c^4} - mc^2$.

The  rescaled   dimensionless  version  (devoid of any physical units) of the ultrarelativistic   operator  $\hat{H} = \sqrt{- \Delta }$, is
often   named the  Cauchy operator  and belongs to the
  $\alpha \in (0,2)$ - family of  fractional  Laplacians (their negatives are called  L\'{e}vy stable generators).
 An investigation of spectral properties of  these  operators  "in the energy landscape  described by a potential $V$",
  see e.g.  \cite{lorinczi,stef} and references there in,
  has a long history. In the  specific context of relativistic  generators   the    eigenvalue problem
  $\hat{H}f = (\hat{H}_m + V)f = Ef, \, m\geq 0$   can be traced back
  to  \cite{weder,herbst,carmona, carmona1}, see also  \cite{lieb,lorinczi1} for further developments on the mathematical  level  of rigor.

The   ultrarelativistic   operator is  nonlocal (quasirelativistic
and $\alpha$-stable operators likewise) and we employ its  integral
definition (involving a  suitable function $f(x)$, with $x\in
\mathbb{R}^d$), that  is  valid in space   dimensions $d\geq 1$,
\cite{ZG1,kwasnicki}: \be
(-\Delta)^{1/2}f(x)=\mathcal{A}_{d}\lim\limits_{\varepsilon\to 0^+}
\int\limits_{\mathbb{R}^d\cap\{|y-x|>\varepsilon\}}
\frac{f(x)-f(y)}{|x-y|^{1+d}}dy. \ee

Eq. (1) derives from the   more  general integral  definition  of the $\alpha \in (0,2)$ -  L\'{e}vy stable  rotationally symetric operator   \cite{ZG1}
and is hereby specialized to $\alpha =1$.   The (L\'{e}vy measure)
normalization coefficient $\mathcal{A}_{d}$ reads
$\mathcal{A}_{d}=\frac{2\,
\Gamma(\frac{1+d}{2})}{\pi^{d/2}|\Gamma(-\frac{1}{2})|}$. We are
primarily  interested in  $d=1$ and $d=3$,  hence
$\mathcal{A}_{1}=\pi^{-1}$   and  $\mathcal{A}_{3}=\pi^{-2}$
respectively. Note that the  integral singularity is  overcome  by
referring to the Cauchy principal value (limiting) procedure.

Our departure point is the  recent paper  \cite{ZG1},  where the employed exterior Dirichlet
 boundary data  were interpreted as  the  infinite spherical well  enclosure   for  the $3D$  ultrarelativistic operator.  Spectral links with the $1D$
 infinite  well problem  (that  actually  has been  solved in \cite{kwasnicki0}, see also  \cite{ZG2,ZG3} and \cite{duo})  have  proved to be instrumental  for the   $3D$
  derivations in Ref. \cite{ZG1}, see also \cite{dyda0,dyda}.

The  finite well enclosure  for operators of the form (1) is introduced by means of
a  finite, explicitly radial  ($r = \sqrt{x^2+y^2+z^2}$),   nonnegative   potential $V$:
\be
V(x,y,z) \equiv V(r) = \left\{
           \begin{array}{ll}
             0, & \hbox{$r  < 1$;} \\
             V_0>0, & \hbox{$r  \geq 1$.}
           \end{array}
         \right.
\ee
One should keep in  mind that no physically relevant  dimensional constants are explicitly involved in the discussion. In this connection, see e.g.
 the appendices in Ref. \cite{GZ} on how to  get rid of them, and   how to  reintroduce them if missing.
We  point out that  the  recalibrated  energy scale is adopted (the bottom of the essential spectrum is shifted from $0$ to $V_0$),
 to conform with the past  analysis of deep $1D$ wells and their spectral convergence towards the  $1D$ infinite well problem,  \cite{ZG2,ZG3}.

Presently, we   address the  $3D$ eigenvalue problem $(\sqrt{-\Delta} + V)f=E f$  with $E>0$  and $f\in L^2(R^3)$, under the
above finite spherical well  (2)  premises.
  Our  major  goal is to  recover the spectral data  (bound states and  respective  eigenvalues)  for the  {\it  shallow}   well.
   In the analysis, we  shall  obtain approximate   eigensolutions  belonging to  purely  radial  $l=0$ and orbitally nontrivial  $l\geq 0$ series.

In Ref. \cite{GZ}  we have found  a  couple of explicit   $1D$  finite well  (deep, but also exemplary shallow cases)
 eigensolutions  for the ultrarelativistic \cite{ZG2} and   quasirelativistic \cite{GZ}  operators.
   We have identified  regularities in the behavior of spectral solutions for:  (i)  the  increasing well depth interpolation  towards the $1D$
   infinite well solution,  (ii) $m \in (0,\infty )$ interpolation   of the quasirelativistic spectrum between the extremities of  the
   ultrarelativistic  and conventional  nonrelativistic spectra.
   Exemplary   well height $V_0$ values were $5, 20, 50, 100, 500$.  The $V_0=500$  well  has been found to  be  spectrally close to the
    infinite well,  while the $V_0=5$  one   could  be  regarded  as shallow.

   In passing let us mention that,  for the exemplary $1D$ well with $V_0=5$,  we have demonstrated the existence of at most $N=3$ bound states \cite{ZG2}.
   In the quasirelativistic case, for $V_0=5$,  we have investigated the $m\in [0.1, 10]$ mass parameter  variability  interval.
    An instructive Table VI in Ref. \cite{GZ} shows that for $V_0=5$, and $m=0.1$
the     quasirelativistic well (and ultrarelativistic likewise) accommodates  more    bound states  (three), than the  corresponding
 nonrelativistic well  (one).

Coming back to the ultrarelativistic case,   in the    $1D$  finite well  at least one  bound  (ground)  state is  known always to exist,
 irrespective of how shallow the well is.
  In the considered presently   $3D$ case, the situation  is different and for too shallow wells the ground state may not exist at all,
    \cite{carmona,carmona1}  and \cite{lorinczi}.   This ground state  existence  issue we shall address by exploiting the link of radial
    eigensolutions of the  $3D$
 finite well    with appropriate eigensolutions of  the $1D$ finite well.

   To our knowledge,  no  explicit existence thresholds   (e.g. the  well height $V_0$  specific choice),  for the existence of the ground state
   or first  excited  states  in the  $3D$ ultrarelativistic shallow well,
     were established as  yet.  As well, the orbital $l\geq 0$  dependence  has never  been investigated  for the $3D$ finite well.
      We attempt to close this gap.    As a byproduct of the  discussion,
      we are capable of  retrieving explicitly  an information about  the  maximal number of bound states,
        given the depth $V_0$ of  the  shallow  spherical well.

{\bf Remark 1:} Since  the ultrarelativistic operators (1)   belong  to the $\alpha \in (0,2)$ - family of  fractional Laplacians
$(-\Delta)^{{\alpha }/2} $, it is useful to mention that under the finite well premises,
 \cite{lorinczi,thanks}:  (i) if  d=1,  the ground state exists for  any $V_0>0$ provided $1\leq \alpha <2$, while
   the   $0<\alpha <1$ case needs the well to be deep enough,  (ii)  if $d=2$
one expects \cite{carmona,carmona1} that,  for every $\alpha \in (0,2)$,   deep enough well is necessary for the ground state formation;
however  it is possible to prove \cite{thanks} the existence  of the ground state for all $V_0>0$ if $1\leq \alpha <2$,
(iii) if $d\geq 3$, then   for  all $\alpha \in (0,2)$, the well needs to be deep enough to accommodate  a ground state.\\
 It is instructive to recall that for the case of the familiar operator  $- \Delta $,  the finite well enclosure is known to yield  the ground state for all $V_0$ if $d\leq 2$,
  while  for $d\geq 3$ the well depth needs to be large enough, \cite{carmona,carmona1}.

{\bf Remark 2:} Concerning the  ground state existence  in a  finite well, for nonlocal Schr\"{o}dinger operators with decaying potentials
\cite{kaleta,thanks} it is possible
 to derive lower bounds on the depth of the potential well in order to have a ground state.
 That in principle comes out  \cite{thanks} from the observation in point  (3) of   Remark 4.1 in  Ref. \cite{kaleta}, p.$27$.
On the other hand, the upper bound on the number of bound states (generalization of the Lieb-Thirring bound) provided by Corollary 2 in
 Ref. \cite{fumio}, can be adopted to the finite well setting.
In  particular, if that bound is less than $1$,  the well is too shallow to allow for the  existence  of a ground state, \cite{thanks}.

{\bf  Remark 3:} In the mathematically oriented  literature on spectral problems for Schr\"{o}dinger-type operators,
\cite{carmona}-\cite{lorinczi1} and \cite{kaleta,fumio,thanks}
and specifically in the context of finite wells,  it is customary  to employ potentials that are  purely  negative
 (or  have  a "substantial"  purely negative part).
 Under such circumstances one obtains a purely negative discrete spectrum, if in existence.
In Eq. (2) we have modified the customary finite well   energy scale.  Typically, in the literature one assumes $V = - v$  with $v>0$ inside the well
 and $0$ at its boundaries and beyond  the well.
If we shift the  energy scale by $v$,   an equivalent spectral problem  arises.
   Indeed,  let us  start from any solution  of  $\sqrt{-\Delta }f  + (V -E)f = 0$.  It is clear that
   $V- E = (V +v) - (E+v)$.
     By choosing  $v=V_0$,  we  obtain   $V- E =  (V +V_0) - (E + V_0)$, which corresponds to the eigenvalue problem with the nonnegative-definite
      finite  well  potential   $(V+V_0)$  of Eq. (2) and    strictly positive eigenvalue $(E+V_0)$.

\section{Existence threshold for the ground state in the $3D$  spherical  well.}

Let  $\mathbf{x}\in \mathbb{R}^3$. From now on we shall simplify the notation and redefine the nonlocal ultrarelativistic operator (1) as follows
\be
Af(\mathbf{x})=\frac{1}{\pi^2}\int\limits_{\mathbb{R}^3}d^3u\frac{f(\mathbf{x})-f(\mathbf{u})}{|\mathbf{x}-\mathbf{u}|^4},
\ee
presuming that whenever a singular integral appears, the Cauchy principal value recipe is enforced.  That extends to (otherwise looking formal, see below) decompositions of a singular integral
 into sums or differences of  singular integrals.  C.f. formulas (4), (5) in  Ref. \cite{ZG1} and note that in the main body of that  paper the  $(p.v.)$  (principal value regularization) symbol
 has been   skipped for notational simplicity. We shall proceed analogously in the present paper.

Let us pass to the ultrarelativistic eigenvalue problem  $(A+V)f=E f$,  where
  $E>0$  and $f\in L^2(R^3)$ is such that the integral (3) exists,  with  $A$  defined  by (3) while  $V$  by (2).

Like in the  infinite well \cite{ZG1},  in the finite well   we look for the ground state in the purely radial form,
 e.g. $f(\mathbf{r})=f(r), \mathbf{r} \in \mathbb{R}^3$.
In view of the presumed radial symmetry, it is natural to pass to  spherical coordinates:
$x=r\cos\phi\sin\theta, y=r\sin\phi\sin\theta, z=r\cos\theta$ where $r\geq 0$, $\theta \in [0,\pi )$ and
 $\phi \in [0,2\pi )$.

Because of the  assumption  $f(\mathbf{r})=f(r)$, the discussion of the eigenvalue problem  may be safely restricted  to  $\mathbf{r}= (0,0,|z|)$ with $z\in \mathbb{R}$.
Accordingly,  we get (remembering about an implicit Cauchy $(p.v)$ regularization)
\be
(Af)(0,0,|z|)=\frac{1}{\pi^2}\calka\frac{(f(|z|)-f(r))r^2\sin\theta}{(r^2+|z|^2-2r|z|\cos\theta)^2}=
\left\{
  \begin{array}{ll}
    \frac{4}{\pi}\int\limits_0^\infty\frac{f(|z|)-f(r)}{r^2}dr, & \hbox{$|z|=0$;} \\
    \frac{1}{\pi |z|}\int\limits_0^\infty (f(|z|)-f(r))r\left(\frac{1}{(r-|z|)^2}-\frac{1}{(r+|z|)^2}\right)dr, & \hbox{$|z|\neq 0$.}
  \end{array}
\right.
\ee
Effectively, $A$  is an integral operator with respect to one variable $r$ only.

Inspired by observations of  Ref. \cite{dyda}, by means of  both
numerical and   analytic arguments,  we have demonstrated in  Ref.
\cite{ZG1},   that  there is an intimate link between
 a subset of spectral solutions for the $1D$ infinite Cauchy well with those of  $3D$  infinite spherical well.  Indeed, radial 3D well
 eigenfunctions can be set in correspondence with   odd (excited)  eigenfunctions of the $1D$ well, while the corresponding  eigenvalues coincide.   This link persists under the  finite well premises.

Indeed, let $|z|\neq 0$.  Because of our assumption $f(\mathbf{r})=f(r)$   about the sought for $3D$  eigenfunction, we have
\be
\left((A+V)f\right)(0,0,|z|)=\left(I_1-I_2\right)+Vf(|z|)=E f(|z|),
\ee
where (the formally looking  subtraction of two singular integrals is carried out in the sense of $(p.v.) (I_1 - I_2)$, compare e.g. Ref. \cite{ZG1}):
\be
I_1=\frac{1}{\pi^2}\left(\calka\frac{f(|z|)r^2\sin\theta}{(r^2-2r|z|\cos\theta+|z|^2)^2}\right)=\frac{f(|z|)}{\pi |z|}\int\limits_0^\infty  r\left(\frac{1}{(r-|z|)^2}-\frac{1}{(r+|z|)^2}\right)dr,
\ee
and
\be
I_2=\frac{1}{\pi^2}\left(\calka\frac{f(r)r^2\sin\theta}{(r^2-2r|z|\cos\theta+|z|^2)^2}\right)=\frac{1}{\pi |z|}\int\limits_0^\infty rf(r)\left(\frac{1}{(r-|z|)^2}-\frac{1}{(r+|z|)^2}\right)dr.
\ee

Let us extend the domain of definition of the function  $f$ from $r
\in \mathbb{R}_+$ to $u  \in \mathbb{R}$.  We demand $f(u)$ to be an
even function, i.e. $f(u)= f(-u)$.  That entails the change of
variables and  allows to modify the range of integrations in the
second integrands of formulas (6) and (7).

 Let us  consider separately cases  $z>0$ and $z<0$. Assuming $z>0$, the parity ansatz allows for a replacement $r \rightarrow u$  in first integrand, while $r\rightarrow -u $ in the  second,
 in both (6) and (7). Accordingly we have
\be
I_1=\frac{f(z)}{\pi z}\int\limits_{-\infty}^\infty  \frac{u}{(u-z)^2}du,
\ee
and
\be
I_2=\frac{1}{\pi z}\int\limits_{-\infty}^\infty  \frac{uf(u)}{(u-z)^2}du.
\ee

Assuming $z<0$ we arrive at the very same formulas, provided  $r$ goes into  $-u$ in the first integrand while $r\rightarrow u$ in the second one.
 Consistently, (6) and (7) take the form (8) and (9) respectively for all  $z\in(-\infty,0)\cup(0,\infty)$.

Accounting for (we recall about the implicit $(p.v)$ recipe)
 \be
\int\limits_{-\infty}^\infty\frac{u}{(u-z)^2}du=\int\limits_{-\infty}^\infty\frac{(u-z)+z}{(u-z)^2}du=
\int\limits_{-\infty}^\infty\frac{z}{(u-z)^2}du,
\ee
we finally give the eigenvalue problem (5) (originally restricted to $z=|z|$) to another,    purely  one-dimensional form  with  $z\in \mathbb{R}$:
\be
\frac{1}{\pi}\left(\int\limits_{-\infty}^\infty\frac{zf(z)}{(u-z)^2}-\int\limits_{-\infty}^\infty\frac{u f(u)}{(u-z)^2}\right)+V z f(z)=E z f(z).
\ee
The left-hand-side of the above eigenvalue formula  is an integral expression ($(p.v)$-regularized) for the $1D$
ultrarelativistic operator (1)  (c.f. also   Eq. (4) in Ref. \cite{ZG1}).

Actually,   Eq. (11) is   the $1D$ eigenvalue problem of the form  $(A+V)g =Eg$  where $g(z)= z f(z)$
 with $f$ being an even function of its argument $z\in \mathbb{R}$.  Thus  $g$ is an odd function.

Accordingly, if an eigensolution  of the $3D$ well spectral  problem  $f(|z|)$  can  extended  to an even  function  $f(z)$ on $\mathbb{R}$, then  $g(z)= z f(z)$ is an odd eigensolution
of the 1D well spectral  problem. Both functions share the same eigenvalue.  Clearly, $g(z)$ must correspond to an excited $1D$ well level.

  The previous reasoning can be inverted and
entails the usage of odd $1D$ well eigensolutions  of the form $z f(z)$  to generate a corresponding family of   purely radial  eigensolutions $f(|z|)$  of the  3D well.
That is paralleled  by  a  spectral  property    $E_{(k,0)}(d=3) = E_{2k}(d=1)$ with $k=1,2,..$,  where  the label  $0$ in $(k,0)$ tells  us  that in $3D$ we generate
 purely radial $l=0$  eigenfunctions,  (see e.g. also \cite{ZG1}).

The above statement  reduces the search for radial eigensolutions in $3D$ well of radius $1$,  to that of identifying   odd eigensolutions in
the $1D$ well of size $(-1,1)$, given a common  positive   $V_0$ value.
Since  in  the  1D case an odd function corresponds to an excited state, we have  at the same time reduced the existence issue for the $3D$ ground state
 in a shallow well to that of the   existence of the first
excited state in the  affiliated   1D  shallow   well.

 We know that in $1D$ well the ground state always exists, but clearly  there is a treshold value for $V_0$
 below which no more bound state (e.g. at least one excited) is  in existence. That is a purely technical
  reason for why in the $3D$ shallow well the ground state may not exist at all, if the well is too shallow.

We shall estimate the threshold value   $V_0$  for the finite well (2), for which a ground state  existence  in   $3D$  will be granted.
 Our discussion (4)-(11) tells that this amounts to the
 existence threshold for the first  excited eigenfucntion for the affiliated  $1D$ finite well.
Based on our previous $1D$ experience, \cite{ZG2}, we shall look for approximate  excited (odd)   eigensolutions of Eq. (11).
 Once  $f(z)$ is determined  we know the the restriction of $f(z)/z$ to  the interval  $(0,1)$ coincides  point-wise with the radial
  eigensolution of the $3D$ finite well problem.

Even in the $1D$ ultrarelativistic case, no analytic expressions for the eigenfunctions or eigenvalues  are known. However it is possible to deduce fairly accurate  approximate expressions by employing
appropriate  numerical methods, \cite{ZG2}.
As in Ref. \cite{ZG2} we shall use the Strang decomposition method, whose basic tenets are outlined for completness.  Its more detailed description can be found in Section II of Ref. \cite{ZG2}.

Let $H=A+V$, c.f. (2) and (3).   To deduce stationary states, we
invoke the (Euclidean looking)  evolution rule    $\exp(-hH)$, $h>0$
which on "short-time-intervals"  $h\ll 1$ can   be given an
approximate form: \be e^{-hH}\simeq e^{-hV/2}e^{-hA}e^{-hV/2}\simeq
e^{-hV/2}(1-hA)e^{-hV/2}=S(h) \ee

The notation convention for the eigenfunctions is
$f_{n,l}^{V_0}(r,\phi,\theta)$ with  $n=1,2,...$, $l=0,1,...$ and
analogously for the  eigenvalues  $E_{n,l}^{V_0}$.  The upper index
needs to  be modified  to indicate the $k$-th step of the coarse-grained
"evolution" algorithm.

   Accordingly, the operation of  $S(h)$  upon the approximate eigenfunction  after  $k$ "evolution" steps reads:
\be
S(h)\phi_{n,l}^{V_0,(k)}(|z|)=e^{-hE_{n,l}^{V_0,(k)}}\phi_{n,l}^{V_0,(k)}(|z|)\sim e^{-hE_{n,l}^{V_0}}\psi_{n,l}^{V_0}(|z|)
\ee
where the corresponding  approximate eigenvalue  after the $k$-th "evolution" step  is given by the expression
\be
E_{n,l}^{V_0,(k)}(h)=-\frac{1}{h}\ln(\mathcal{E}_{n,l}^{V_0,(k)}(h)),\qquad \mathcal{E}_{n,l}^{V_0,(k)}(h)=<\phi_{n,l}^{V_0,(k)}|\psi_{n,l}^{V_0,(k+1)}>=<\phi_{n,l}^{V_0,(k)}|S(h)\phi_{n,l}^{V_0,(k)}>.
\ee
where $<\cdot|\cdot>$  indicates the conventional $L^2$ scalar product.\\

At this point  we are inspired by the  the observation of Ref. \cite{ZG1}, and  irrespective of the well depth/height,   we  can  interpret
  the  finite  spherical well   spectrum   to have the  form   of an ordered set of strictly positive eigenvalues, that naturally splits
   into non-overlapping,   orbitally labeled   $E_{(n,l)}$ series.  Consistently, for each fixed  value of $l=0,1,2...$, the  label  $n=1,2,...$
    enumerates     consecutive eigenvalues  within the  particular  $l$-th series.

{\bf Remark  4:} \\
(i) The definition of  $A$,    Eq. (3),  introduces the  integral
that  needs to be  evaluated numerically.  Its  value depends on the
choice of  integration intervals and their partitioning. The finer
partitioning results in more accurate approximations,
the price paid is quickly growing computation time.
Based on our previous experience   \cite{ZG2}  we consider  the partition unit   $0.001$ to be optimal for our purposes.\\
(ii)  Extending integration intervals to infinity is beyond the reach  of simulation preocedures.  Therefore one must decide about an optimal
 (not too large)  integration interval $(-a,a)$ in $\mathbb{R}$, c.f. \cite{ZG2}. \\
(iii)  The latter $(-a,a)$ limitation automatically involves computational  problems to be  taken care of,  since  e.g.   (10)  is no longer literally  valid:
\be
\int\limits_{-a}^a\frac{r}{(r-z)^2}dr=\int\limits_{-a}^a\frac{z}{(r-z)^2}dr+\ln\left|\frac{a-z}{a+z}\right|.
\ee
It is clear  that the  $a \rightarrow  \infty $ limit, valid for all $z\in \mathbb{R}$, would restore (10).\\

{\bf Remark 5:}
 Coming back to  (11), while being solved approximately,  we note that for a fixed integration interval $(-a,a)$, we can still   optimize  (in fact increase)  the accuracy of the
 eigenvalue computation,  provided $a$ is not too small.  Indeed,   let   $f(x)$   be a $1D$ approximate eigenfunction of  $H$, where   $[-a,a]$ is the integration interval.
 We have:
 \be
H(a)f(x)\sim [(-\Delta)^{1/2}+V(x)]f(x)=\frac{1}{\pi}\int_{|z|\leqslant a}\frac{f(x)-f(x+z)}{z^2}dz+V(x)f(x)\sim E(a)f(x). \nonumber
\ee
Let  $g(x)$  be an approximate eigenfunction evaluated with the choice  $[-b,b]$, $b>a$ of the integration interval. Then:
\be
\begin{split}
H(b)g(x)\sim \frac{1}{\pi} &\int_{|z|\leqslant b}\frac{g(x)-g(x+z)}{z^2}dz+V(x)g(x)\sim E(a)g(x)+\frac{1}{\pi}\int_{a\leqslant |z|\leqslant b}\frac{g(x)-g(x+z)}{z^2}dz\sim\\
&E(a)g(x)+\frac{2}{\pi}g(x)\int_a^b\frac{dz}{z^2}\sim\left[E(a)+\frac{2}{\pi}\left(\frac{1}{a}-\frac{1}{b}\right)\right]g(x)\sim E(b)g(x)  \nonumber
\end{split}
\ee
Accordingly, for sufficiently large values of $a$ and $b$ we have   $ E(b)-E(a)\sim\frac{2}{\pi}\left(\frac{1}{a}-\frac{1}{b}\right) $.  We can thus estimate a difference between the eigen computations involving different (increasing) intervals.
Namely,  $\frac{2}{\pi}\left(\frac{1}{50}-\frac{1}{100}\right)\sim 0.0064$, next  $\frac{2}{\pi}\left(\frac{1}{100}-\frac{1}{200}\right)\sim 0.0032$,
$\frac{2}{\pi}\left(\frac{1}{200}-\frac{1}{500}\right)\sim 0.0019$  and ultimately $\frac{2}{\pi}\left(\frac{1}{500}-\frac{1}{\infty}\right)\sim 0.0013$.
We have checked the validity  and usefulness  of these "interval size  renormalization"   by exemplary simulations, see also \cite{ZG2}.  \\

The threshold value  $V_0>0$ which yields  the ground state existence
in $3D$   actually  comes out as value  for which the $1D$ finite
well has exactly two  bound states.

Executing the  "evolution"  (12)-(14)  numerically, with the initial
data chosen as trigonometric functions (c.f.  (16) in  \cite{ZG1}),
we can readily check that for $V_0=2$  there is no   bound (e.g.  ground) state in
$3D$.

To the contrary, an explicit computation for  $V_0=2.1$  proves that
the $3D$  ground state does exist.  Accordingly, we know for sure
that in the interval $(2,2.1)$   there  exists  $V_0$ for which the
ground state appears, being absent below this value.

   The interval width $0.1$ might look excessively large. However this  is not the case, as
the subsequent discussion will reveal.  (Let us mention that this
localization width for $V_0$  may be made finer, because the
numerical simulations accuracy  can be significantly improved.)

 In Table I we collect the approximate ground state eigenvalues in the $3D$ well with $V_0=2.1$, each obtained for another choice of the
  integration interval $(-a,a)$. We  display the  direct  simulation outcomes  for $a\leq 500$. The last line ($a= \infty $)  contains
  an   eigenvalue  estimate for $a\rightarrow \infty $, i.e. $E_{500}$ that is  "renormalized" by the   missing tail  contribution $0.0013$
  (c.f. Remark 5).
  \begin{table}[h]
\begin{center}
\begin{tabular}{|c||c|}
\hline
a & E ($V_0=2.1$) \\
\hline
\hline
50 & 2.02603\\
\hline
100 & 2.03242\\
\hline
200 & 2.03562\\
\hline
500 & 2.03752\\
\hline
$\infty$ & 2.03882\\
\hline
\end{tabular}
\end{center}
\caption{The ground state eigenvalue in $3D$ well (equivalently, the
second $1D$ well eigenvalue): dependence on  $a$.}
\end{table}

We have performed a simulation procedure (12)-(14) for  $V_0=2$  and $a=50$,  with an outcome  suggesting the existence of
 the bound state with  an approximate  eigenvalue   equal   $1.9926<2$.  However, on the basis of
Remark 5 and \cite{ZG2}  we know that the   simulation outcome  (e.g. the computed eigenvalue)  necessarily  grows  up with the growth of $a$.
 In fact (c.f. Remark 5), for a "true"  eigenstate   we    expect  the missing tail  contribution  to be    $0.0128$.
But then, the   pertinent candidate bound state  eigenvalue would exceed the potential height  $V_0 = 2$, being equal $2.0054$.
That is untenable,  hence we conclude that  for   $V_0=2 $  the   $3D$ ground state    does not exist.

For $V_0=2.1$ the eigenvalues in Table I are definitely  smaller than $2.1$.   We have under  control  an impact of   numerical errors in the
 employed algorithm to be less  than $1\%$  of the computed eigenvalue.   The "renormalization"  by $0.0013$  is harmless as well.
    We  thus  conclude  that  the existence threshold for the $3D$ well ground state is located in the interval   $V_0\in (2,2.1]$.

\begin{figure}[h]
\begin{center}
\centering
\includegraphics[width=80mm,height=80mm]{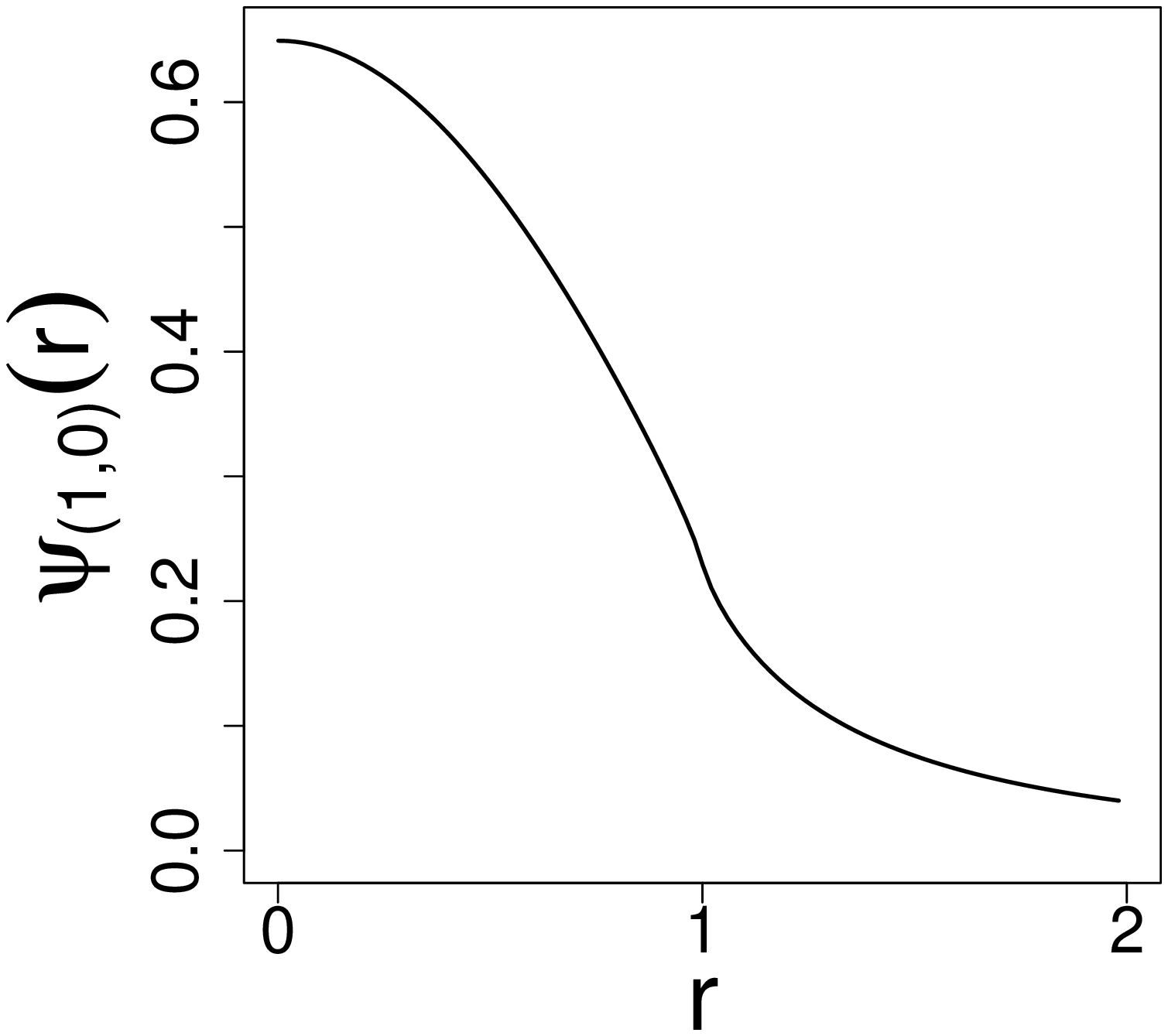}
\includegraphics[width=80mm,height=80mm]{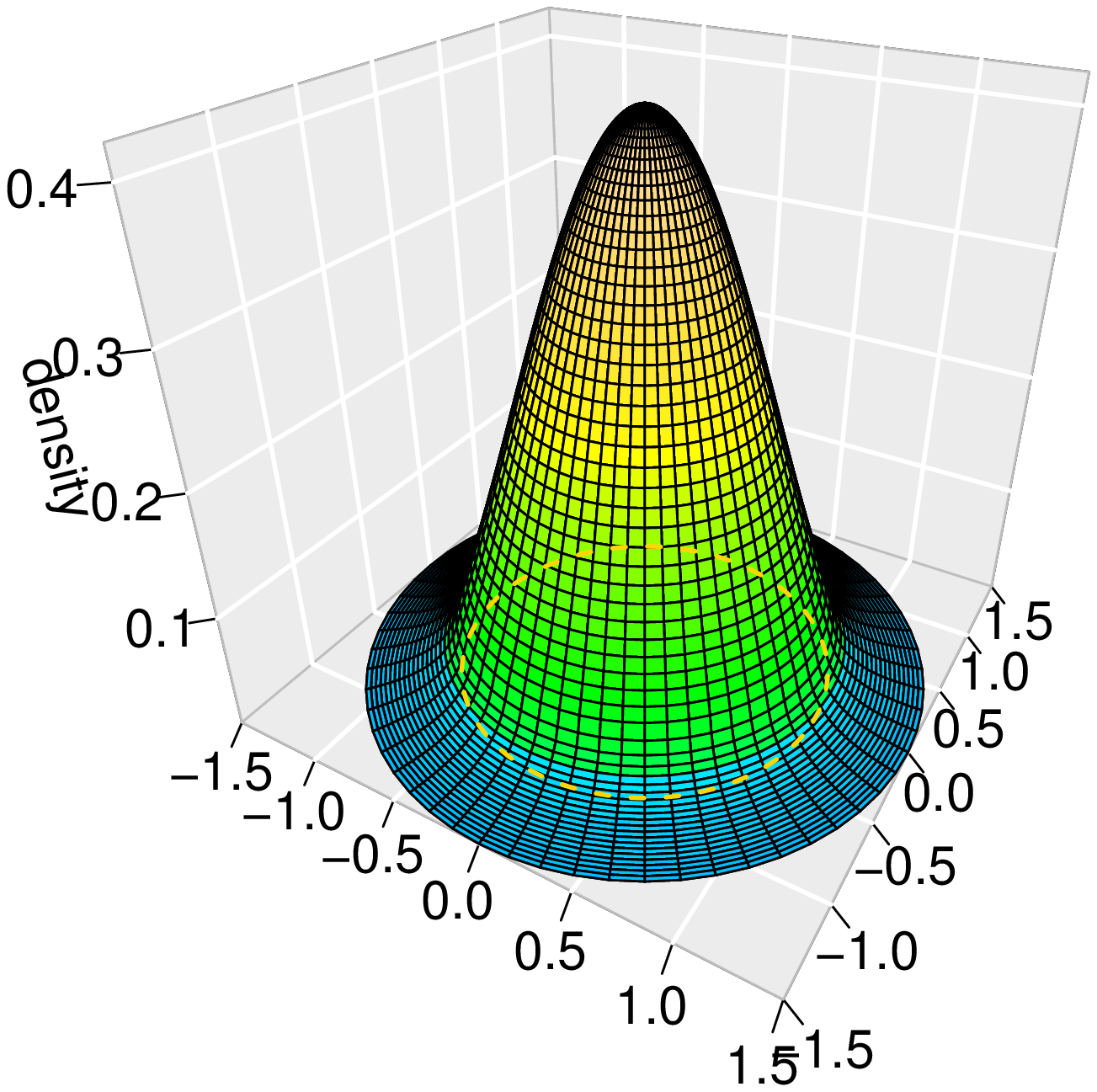}
\caption{The radial ground state and the  probability density for   $V_0=2.1$.  The circle on the   polar   plot
indicates the well boundary $r=1$. The conspicuous tails depict qualitatively  the  tunelling effect.}
\end{center}
\end{figure}

Simulation data allow  us to  deduce an approximate form of the eigenfunction, see e.g. Fig. 1.   As a byproduct of simulations we have computed
the ratio of  probabilities: that of   localizing  the state within the well area  $P_{in}$,  and that referring to the delocalized tail (beyond
 the well) $P_{out}$: .
\be
\frac{P_{out}}{P_{in}}=\frac{\int\limits_1^a r^2\psi^2(r)dr}{\int\limits_0^1 r^2\psi^2(r)dr}=
\frac{1-4\pi \int\limits_0^1 r^2\psi^2(r)dr}{4\pi\int\limits_0^1 r^2\psi^2(r)dr}.
\ee
A concrete numerical outcome depends on $a$, but appears to stabilize for large values of $a$. In particular, for  $V_0=2.1$ and  $a=500$
 we arrive at the ratio value  $0.438$ indicating  a conspicuous strength of the tunneling effect.

\section{Existence thresholds for higher  purely  radial eigenfunctions.}

We can extend the methodology of Section II   to  set existence thresholds for  higher radial bound states in $3D$.
However, before proceeding further with the ultrarelativistic case, let us recall known facts about  the standard  ($(-\Delta + V)f=Ef$)
Schr\"{o}dinger  spectral problem for  the  finite well, \cite{Gr,VO}.

   If energy is measured in units $\hbar ^2/2m$, while any a priori chosen radius  $R>0$ stands for the length unit,  the  dimensional
    no-ground-state-in existence criterion  $\sqrt{\frac{2mV_0R^2}{\hbar^2}}<\frac{\pi}{2}$   takes the form  $V_0< \pi ^2/4\sim 2.45 $.
     We have exactly  $n\geq 1$  bound states if the well potential obeys  inequalities  $(2n-1)^2 \frac{\pi ^2}{4}<  V_0 < (2n+1)^2 \frac{\pi ^2}{4}$.
       The corresponding eigenvalues  are accessible  by means of   numerical methods  only.
 W note  a conspicuous (albeit  rough)  $n^2$ scaling of  consecutive  threshold  values for $V_0$.

 The situation is different in the ultrarelativistic case, where the analogous scaling is  approximately   linear in $n$,  see e.g.  \cite{stef,kwasnicki0}
  for the  related  $1D$ discussion  as well as for  that on   limits of its  $3D$  validity, \cite{ZG1}.

In view of the previously established $1D$-$3D$ spectral link, if we are interested  in the existence of the second and third $3D$ finite  well eigenvalue,
 actually  we need to deduce  the  existence  threshold $V_0$  (we keep the $0.1$ accuracy limitation) for the $4$th and $6$th   $1D$ well eigenvalues.
This we have done numerically with $a=50$. Computation outcomes  are collected in table II, where the notation  $E_{(k,0)}$ with $k=1,2,...$
explicitly introduces the $l=0$ orbital label, \cite{ZG1}.  The pertinent eigenfunctions are purely radial.
\begin{table}[h]
\begin{center}
\begin{tabular}{|c||c|c|c|}
\hline
a & $E_{(1,0)}$ ($V_0=2.1$) & $E_{(2,0)}$ ($V_0=5.2$) & $E_{(3,0)}$ ($V_0=8.3$) \\
\hline
\hline
50 & 2.02603 & 5.13346 & 8.26733 \\
\hline
$\infty$ & 2.03882 & 5.14626 & 8.28013 \\
\hline
\end{tabular}
\end{center}
\caption{$3D$ spherical  well:   approximate   $V_0$ threshold values for the existence of
(i) the ground state, (ii)  first and   (iii) second
excited (purely radial) eigenstates  (first three eigenvalues in the $l=0$ series).
 The cumulative correction of Remark 5, taking $E_{50}$ into $E_{\infty }$,  equals   $0.0128$.}
\end{table}

For each of the considered threshold values, a  lowering of a given  $V_0$ value   by $0.1$  is sufficient for the pertinent bound
 state (i)-(iii) in Table II {\it  not}  to exist.   Some caution is necessary in connection with our   explicit  threshold values $V_0$.
The $(-a,a)$  integrations are carried out  numerically upon a definite partition unit choice (we have set the partition finesse at  0.001,
 c.f. Remark 4).
Further tuning of the  partition finesse  would increase an integration accuracy and   then a residual  modification of  $5.2$ or $ 8.3$
 threshold values might in principle  be necessary on the level of not displayed decimal digits.

\begin{figure}[h]
\begin{center}
\centering
\includegraphics[width=80mm,height=80mm]{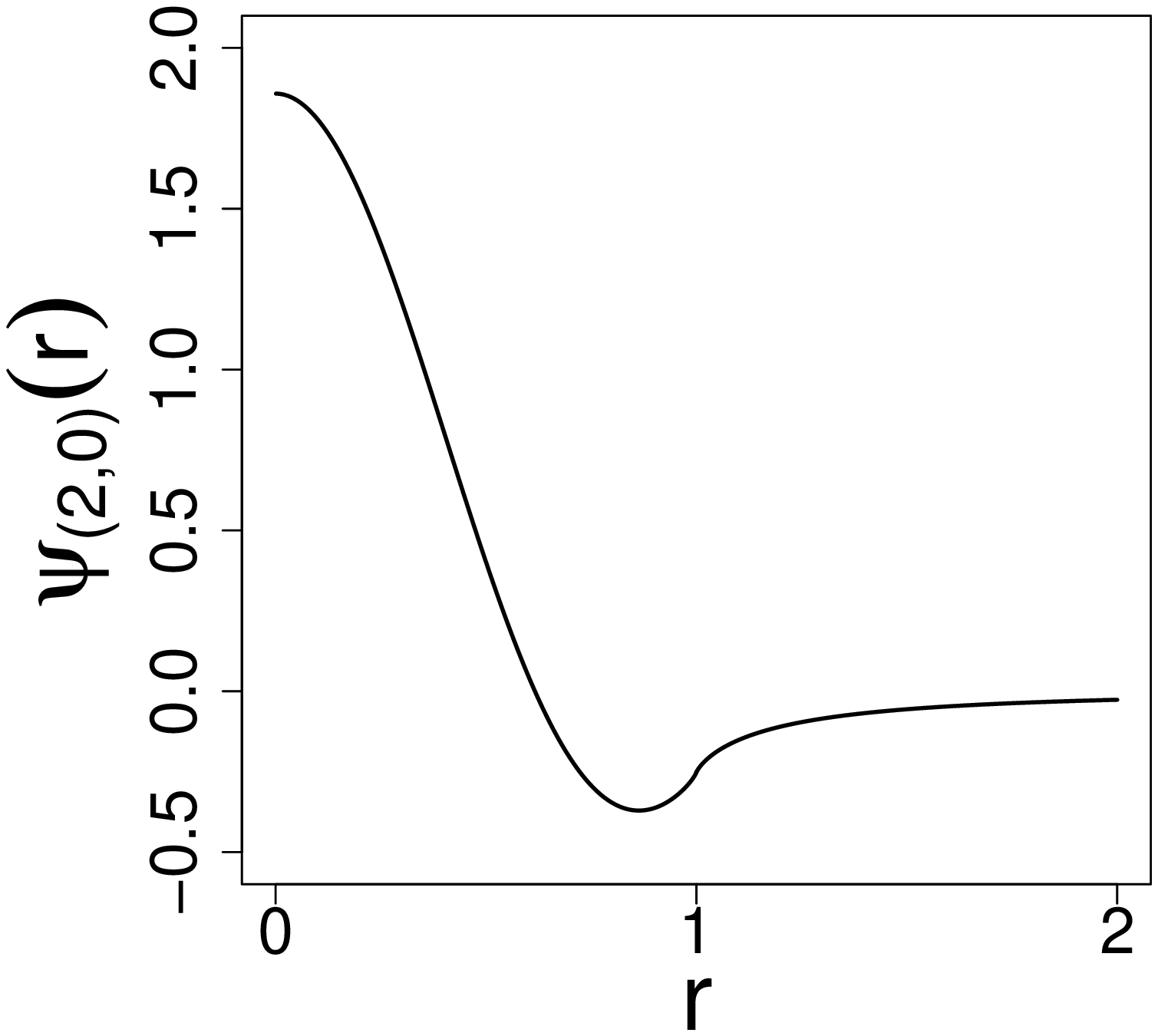}
\includegraphics[width=80mm,height=80mm]{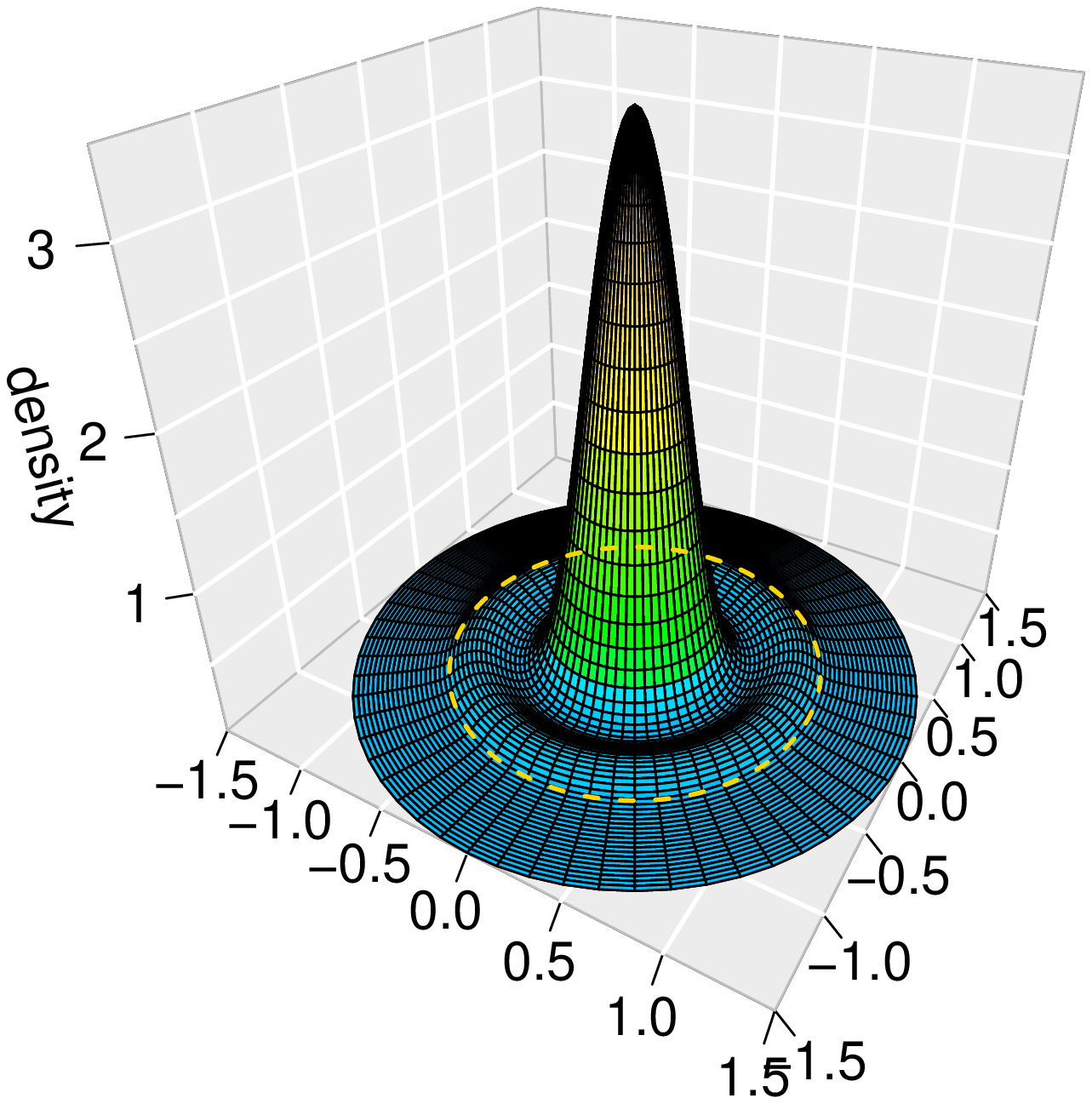}
\caption{Second (first excited)  radial eigenfunction in the $3D$ well,
 together with the  polar plot for  the   related probability density,  for  $V_0=5.2$ and  $a=50$.}
\end{center}
\end{figure}
The related eigenfunction has one nodal set (circle) and   quickly decays (drops down) beyond  the well area.
We add that for  $V_0=5.2$ the ground state energy  ($a=50$)  reads   $2.38033$, to be set against the excited radial state eigenvalue  $5.13346$.\\

The third radial bound state in the $3D$ well does exist for  $V_0=8.3$  and has two  disjoint  nodal sets (circles).

\begin{figure}[h]
\begin{center}
\centering
\includegraphics[width=80mm,height=80mm]{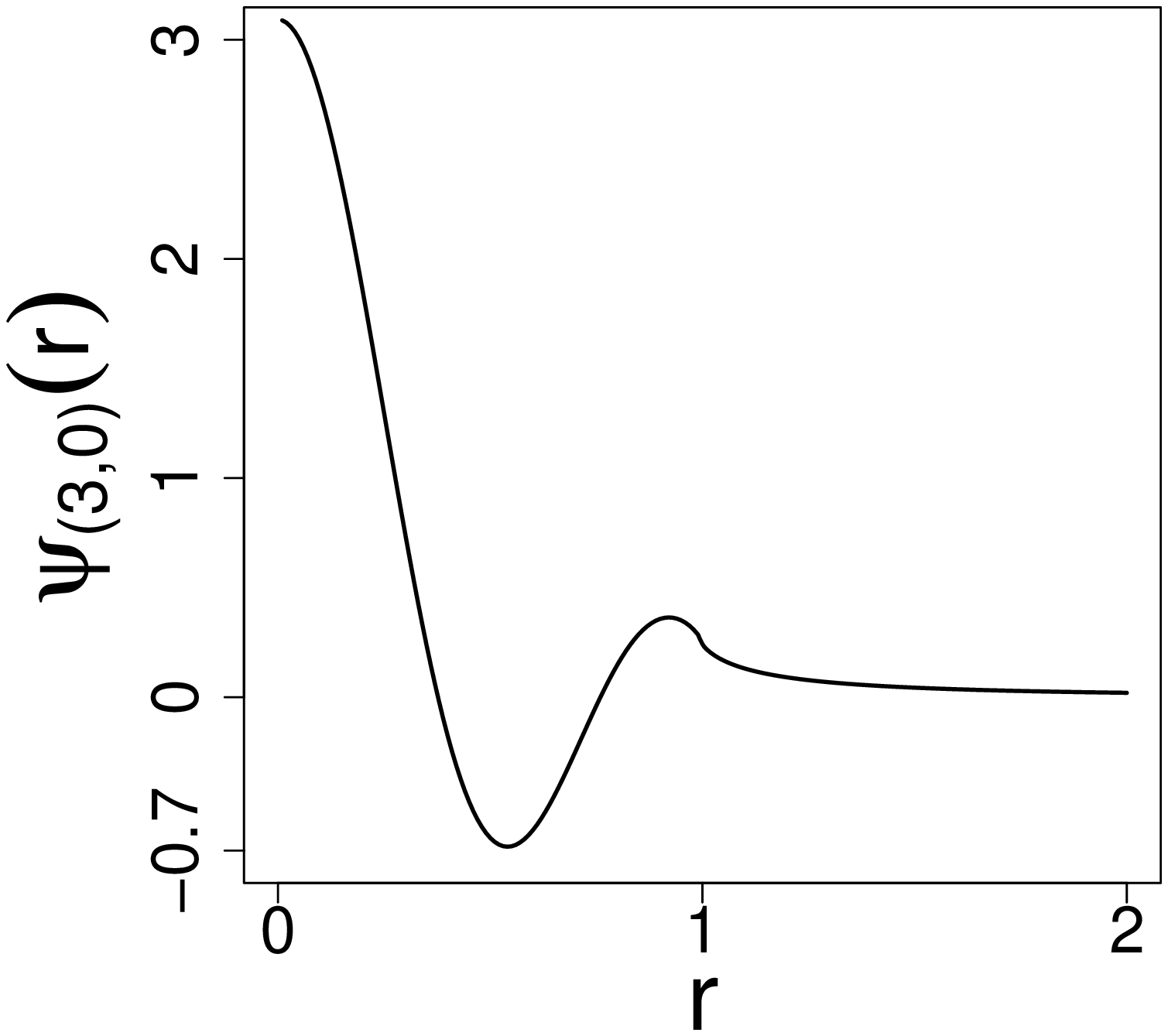}
\includegraphics[width=80mm,height=80mm]{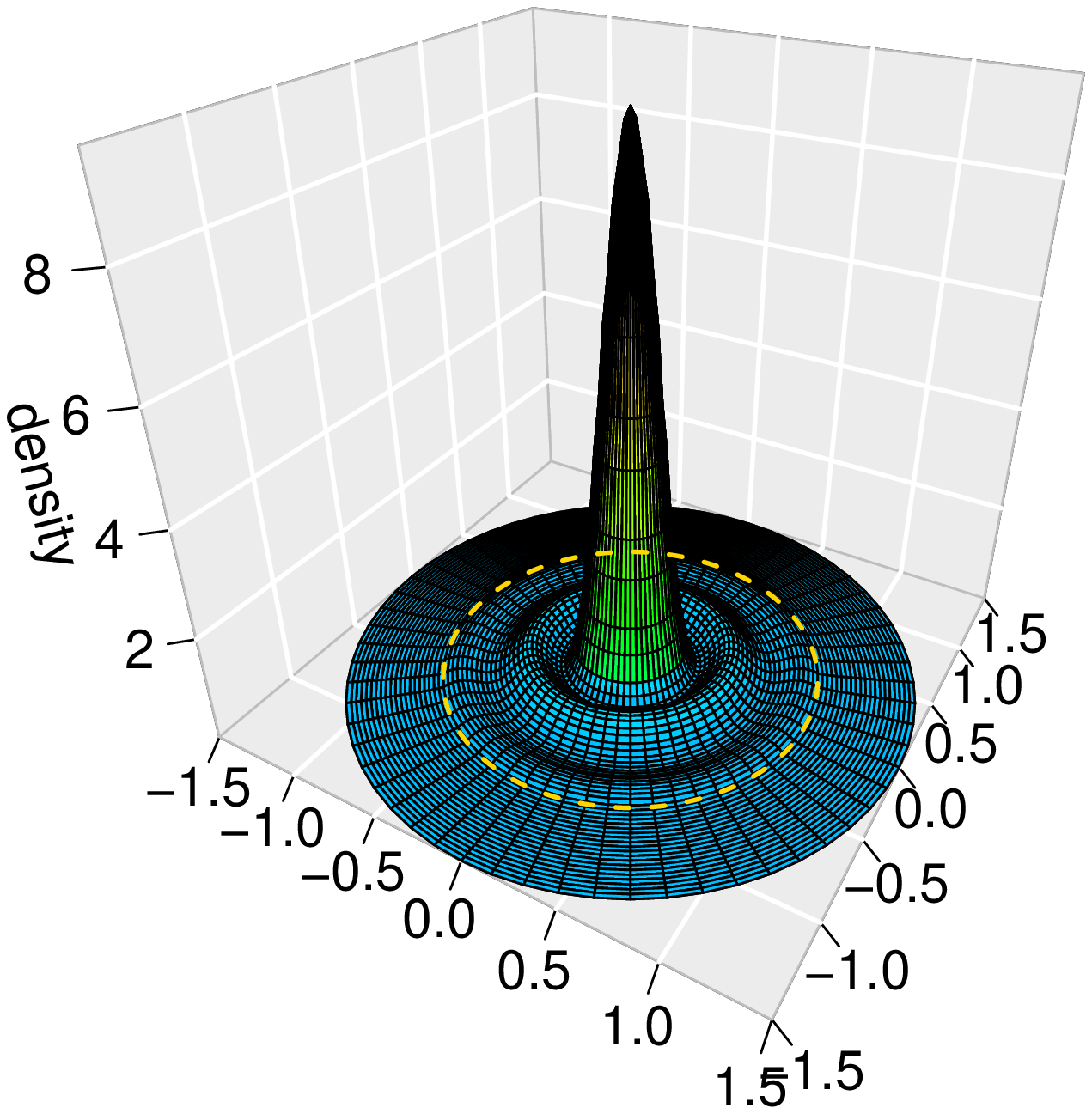}
\caption{Third radial eigenfunction and the  inferred probability density  plot  for  $V_0=8.3$ and $a=50$.
 The circle  on the polar plot indicates the well boundary $r=1$.}
\end{center}
\end{figure}

On the basis of  data presented  in  Table II we realize  that consecutive bound states are allowed to appear if  $V_0$  changes approximately by
  $\pi$  (actually, we get $\sim 3.1$).
 It is consistent with the above mentioned linear  scaling of threshold values, whose sharper  version has been established  for the $1D$
 infinite ultrarelativistic well, \cite{kwasnicki0}. Indeed,  in $1D$  there holds (which is a fairly  good estimate beginning from $n=5$):
\be
E_n=\frac{n\pi}{2}-\frac{\pi}{8}+\mathcal{O}\left(\frac{1}{n}\right),
\ee
so that  $E_{2n+2}-E_{2n}\sim\pi$. Since  for radial $3D$ bound states  the $1D$     eigenvalue
$E_{2n}$  actually  coincides with the $n$th  $3D$ eigenvalue.

\section{Existence thresholds for orbitally nontrivial ($l>0$) eigenfunctions.}

\subsection{$l=1$ eigenfunctions}

In our study \cite{ZG1} of the  infinite spherical well spectral problem,  in addition to purely radial eigenfunctions,   we have identified eigenfunctions that are not purely radial and thence
 belong to   orbital sectors  labeled by  $l\geq 1$.
These observations provide  a   useful  guidance in case of the finite spherical well.  Our further analysis will in part rely on the analytic approach to the   evaluation of  (singular)  integrals, c.f.
sections IV.A and B in Ref. \cite{ZG1}.  Albeit  the ultimate  simulation algorithm    will be  entirely different from that employed in \cite{ZG1}.
 By integrating out all angular contributions, we arrive at purely radial integrals.  The   numerically-assisted  (approximate) solution of the eigenvalue problem, can be addressed by means of the
  so-called Strang splitting  method, succesfully used in $1D$ considerations of Ref.  \cite{ZG2}.  Its outline has been given in Section II.

In Ref. \cite{ZG1} we have shown that the ultrarelativistic  infinite  well eigenfunctions have a generic form
$\psi_{(k,l,m)}(r,\theta,\phi)=f_k(r)Y_{l,m}(\theta,\phi)$,  where
 $Y_{l,m}(\theta,\phi)$  are spherical harmonics in  $\mathbb{R}^3$  and $k=1,2,... $ labels  are uncorrelated with the $l=0,1.2,...$ orbital labels,
  while $|m|\leq l$ for each $l$.
 In the  pertinent  infinite well regime, we have imposed specific requirements concerning the functional form of $f_k(r)$.
 None of them is in use presently, in the finite well  setting. Therefore,  we shall  employ another computation
 method (Strang splitting instead of Mathematica routines) than that of Ref. \cite{ZG2}, while leaving  intact  the factorization
 ansatz $\psi (r,\theta,\phi)=f(r)\,  Y(\theta,\phi) $.

Let  $\textbf{p}=(x_1,x_2,x_3)\in \mathbb{R}^3$. Anticipating the $l=1$  ultimate  eigensolution   we  first look for an
 eigenfunction  in the functional form  $\psi(\textbf{p})=x_3f(p)$,  where  $p=\sqrt{\iksy}$ and   $x_3=r \cos \theta \neq 0$.
The purely radial function  $f(p)$  is at the moment unknown and should follow from the eigenvalue equation
\be
A\psi(\textbf{p})+V(\textbf{p})\psi(\textbf{p})=    (I_1 -I_2)\psi(\textbf{p})  + V(\textbf{p})\psi(\textbf{p}) =  E \psi(\textbf{p}).
\ee

The integral operator   $A$  is here re-defined as a $(p.v.)$ computable  difference of singular
 integrals  $(I_1-I_2)\psi(\textbf{p})$, where
\be
A\psi(\textbf{p})={\frac{1}{\pi^2}} \left( \int\limits_{\mathbb{R}^3} {\frac{\psi(\textbf{p})du}{(\bf{u} - \bf{p})^4}}
-\int\limits_{\mathbb{R}^3} {\frac{\psi(\textbf{u})du}{(\bf{u} - \bf{p})^4}} \right),
\ee
and $\textbf{u}=(u_1,u_2,u_3)$ and $du$ indicates  a three-dimensional integration.

In Ref. \cite{ZG1} we have described how to reduce  $\mathbb{R}$ integrals,   involved in (18) via (1), to the purely radial integration.
C.f. Section IV.A, Eqs. (27)-(36)   there in.  We have constructed a rotation  matrix $S$  in $\mathbb{R}^3$  such that
\be
u_i=s_{i1}v_1+s_{i2}v_2+s_{i3}v_3,\qquad i=1,2,3,
\ee
where  $s_{ij}$ are matrix elements of $S$  such that
\begin{eqnarray}
\sqrt{u_1^2+u_2^2+u_3^2}\xrightarrow{S} \sqrt{v_1^2+v_2^2+v_3^2} = r ,\\
(u_1-x_1)^2+(u_2-x_2)^2+(u_3-x_3)^2\xrightarrow{S} v_1^2+v_2^2+(v_3-\sqrt{\iksy})^2.
\end{eqnarray}
We denote $\textbf{r} = (v_1,v_2,v_3)$ and $|\textbf{v}| = r$.

Keeping in mind that both $I_1$ and  $I_2$  are singular integrals and that  the it is the  $(p.v.)(I_1-I_2)$  recipe  that  removes the involved  all obstacles, we shall
evaluate the integral entries separately, while passing to spherical coordinates with:
\be
\begin{split}
I_1(\bf{p}) &=\frac{1}{\pi^2}\int\limits_{\mathbb{R}^3}\frac{x_3f(p)du}{((u_1-x_1)^2+(u_2-x_2)^2+(u_3-x_3)^2)^2}=
\frac{x_3f(p)}{\pi^2}\int\limits_{\mathbb{R}^3}\frac{dv}{(v_1^2+v_2^2+(v_3-p)^2)^2}\\
&=\frac{x_3f(p)}{\pi^2}\calka \frac{r^2\sin\theta}{(r^2+p^2-2r p\cos\theta)^2}=
\frac{x_3f(p)}{\pi p}\int\limits_0^\infty r\left(\frac{1}{(r-p)^2}-\frac{1}{(r+p)^2}\right)dr,
\end{split}
\end{equation}
and
\be
\begin{split}
I_2(\bf{p})  &=\frac{1}{\pi^2}\int\limits_{\mathbb{R}^3}\frac{u_3f(u)d^3u}{((u_1-x_1)^2+(u_2-x_2)^2+(u_3-x_3)^2)^2}=
\frac{s_{33}}{\pi^2}\calka \frac{r^3f(r)\sin\theta\cos\theta}{(r^2+p^2-2rp\cos\theta)^2}\\
&=\frac{s_{33}}{2\pi p^2}(I_{21}+I_{22})=\frac{x_3}{2\pi p^3}(I_{21}+I_{22})
\end{split}
\end{equation}
where
\be
\begin{split}
I_{21}(p) &=\int\limits_0^\infty r(r^2+p^2)f(r)\left(\frac{1}{(r-p)^2}-\frac{1}{(r+p)^2}\right)dr,\\
I_{22} (p)&=\int\limits_0^\infty rf(r)\left(\ln(r-p)^2-\ln(r+p)^2\right)dr.
\end{split}
\ee

Accordingly,  the $x_3\neq 0$ factor becomes irrelevant and we reduce  Eq. (19) to the form:
\be
(Af)(p)+V(p)f(p)=Ef(p)
\ee
where $(Af)(p)$ is a purely radial integral
\be
(Af)(p)=\frac{1}{2\pi}\int\limits_0^\infty \left[\frac{r}{p}\left(2f(p)-\frac{r^2+p^2}{p^2}f(r)\right)\left(\frac{1}{(r-p)^2}-\frac{1}{(r+p)^2}\right)-\frac{r f(r)}{p^3}\left(\ln(r-p)^2-\ln(r+p)^2\right)\right]dr.
\ee
It is the eigenvalue problem  with respect to $f(p)$ and $E$ which   we shall address by means of the Strang method  of Ref. \cite{ZG2}.
One needs to remember about the $L^2$ scalar product and norm input in the Strang method.
 The scalar product  $<f_1|f_2>$ we  directly infer from $<\psi _1|\psi _2>$   (remembering that $\psi = x_3\, f$):
\be
\calka (r^2\sin\theta) (r\cos\theta f_1(r))(r\cos\theta f_2(r))=\frac{4\pi}{3}\int\limits_0^\infty  r^4f_1(r)f_2(r)dr.
\ee
 Consequently, the normalization coefficient is  given by:
\be
C^2\calka r^2\sin\theta(r\cos\theta f(r))^2=\frac{4\pi C^2}{3}\int\limits_0^\infty r^4f^2(r)dr=1.
\ee

Eigenvalue problems of the type (26) and  (27) have never been studied in the literature. Previously, \cite{ZG1}  we have
addressed that issue for the infinite spherical well. Now the considered spherical  well is  not merely  finite, but   shallow.

The Strang method  has been adopted to solve (approximately)  the pertinent eigenvalue problem in the orbital $l=1$ sector,
c.f. for comparison our infinite well data of Ref. \cite{ZG1}.
We stress that computation outcomes rely both on the choice of then integration boundary $a>0$ and the partition finesse, which we have set at
the value $\Delta x=0.001$.

We have identified the existence of the orbital eigenstate in the $V_0=3.5$ well. We have also verified that for
   $V_0=3.4$  there is no  $l=1$ eigenfunction.

\begin{table}[h]
\begin{center}
\begin{tabular}{|c||c|}
\hline
a & E ($V_0=3.5$) \\
\hline
\hline
50 & 3.43477\\
\hline
100 & 3.44755\\
\hline
200 & 3.45393\\
\hline
500 & 3.45776\\
\hline
$\infty$ & 3.46036\\
\hline
\end{tabular}
\end{center}
\caption{The first (and  the only)  orbital ($l=1$) eigenvalue for  $V_0=3.5$:  the $a$-dependence.}
\end{table}

 We point out that spacings between eigenvalues obtained for different  choices   of  $a$  stay in a conspicuous agreement with our discussion  in Section II.
 These are respectively  $0.01278, 0.00638, 0.00383$  and roughly coincide with doubled values reported in Remark 5.
  We recall that the pertinent discussion has elucidated properties of the $1D$ integrations within the interval $[-a,a]$. In $3D$ we
   integrate with respect to the radial variable $r$, on the interval  $[0,a]$.

The above observation allows in principle to interpolate results obtained for a given value of $a$ towards $\infty $,
   e.g. the $a=500$ eigenvalue might in principle be additively renormalized by
 $2 \cdot 0.0013$.  Such "corrected eigenvalue would read  $E_{(1,1)}^{(V_0=3.5)}=3.46036<3.5$  and is actually  listed in Table III.

In the computation process  for $V_0=3.5$, we  recover the data necessary to depict the radial part of the eigenfunction  $\psi_{(1,1,0)}$,   (that is simply  $rf(r)$)
and the plots of related probability densities   $|\psi_{(1,1,0)}(r,\theta,\phi)|^2$  and  $|\psi_{(1,1,\pm 1)}(r,\theta,\phi)|^2$.
We remind that the probability densities  are $\phi $-independent and polar plots give an accurate visualization of the  spatial properties of eigenfunctions.

\begin{figure}[h]
\begin{center}
\centering
\includegraphics[width=55mm,height=55mm]{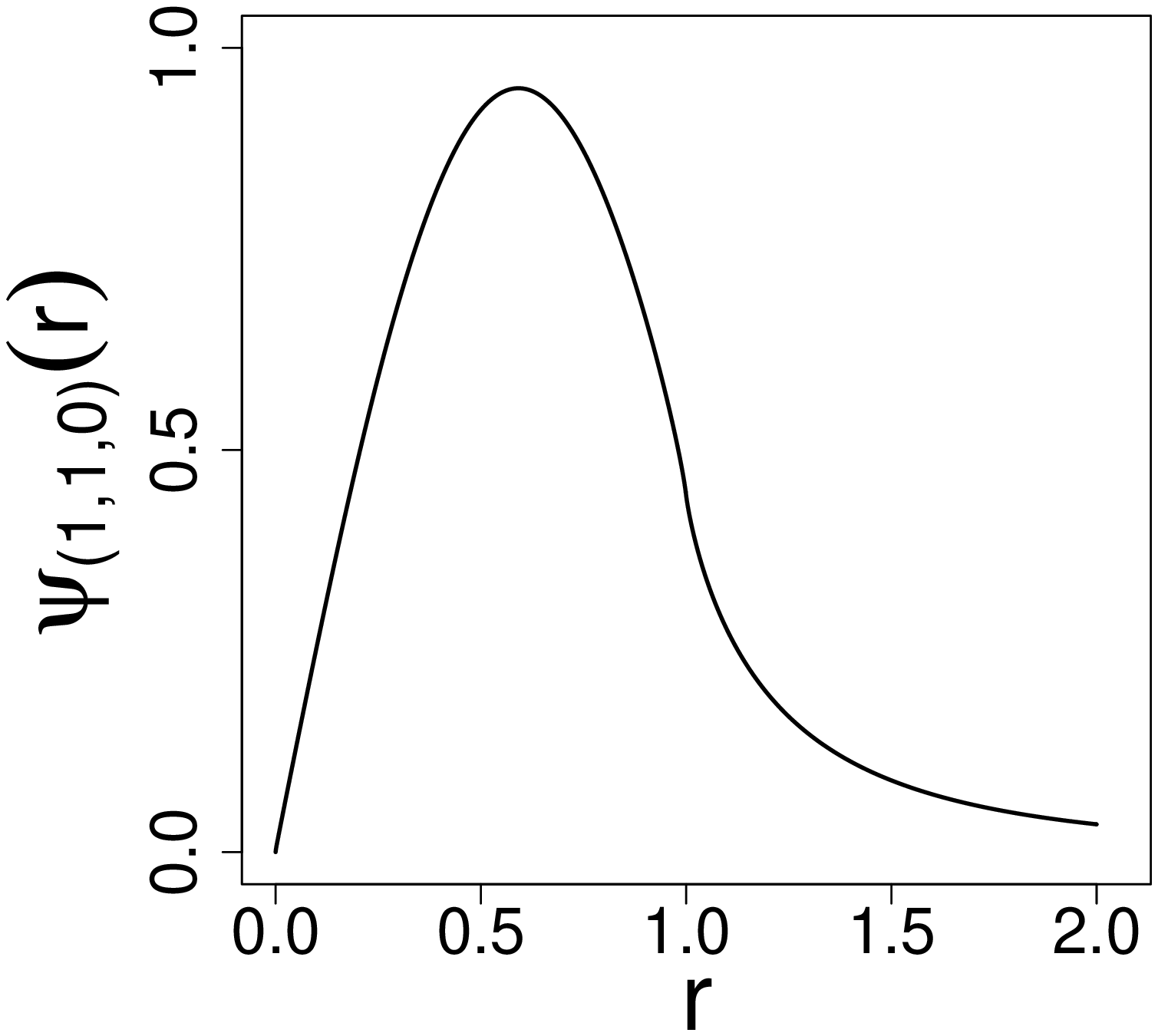}
\includegraphics[width=55mm,height=55mm]{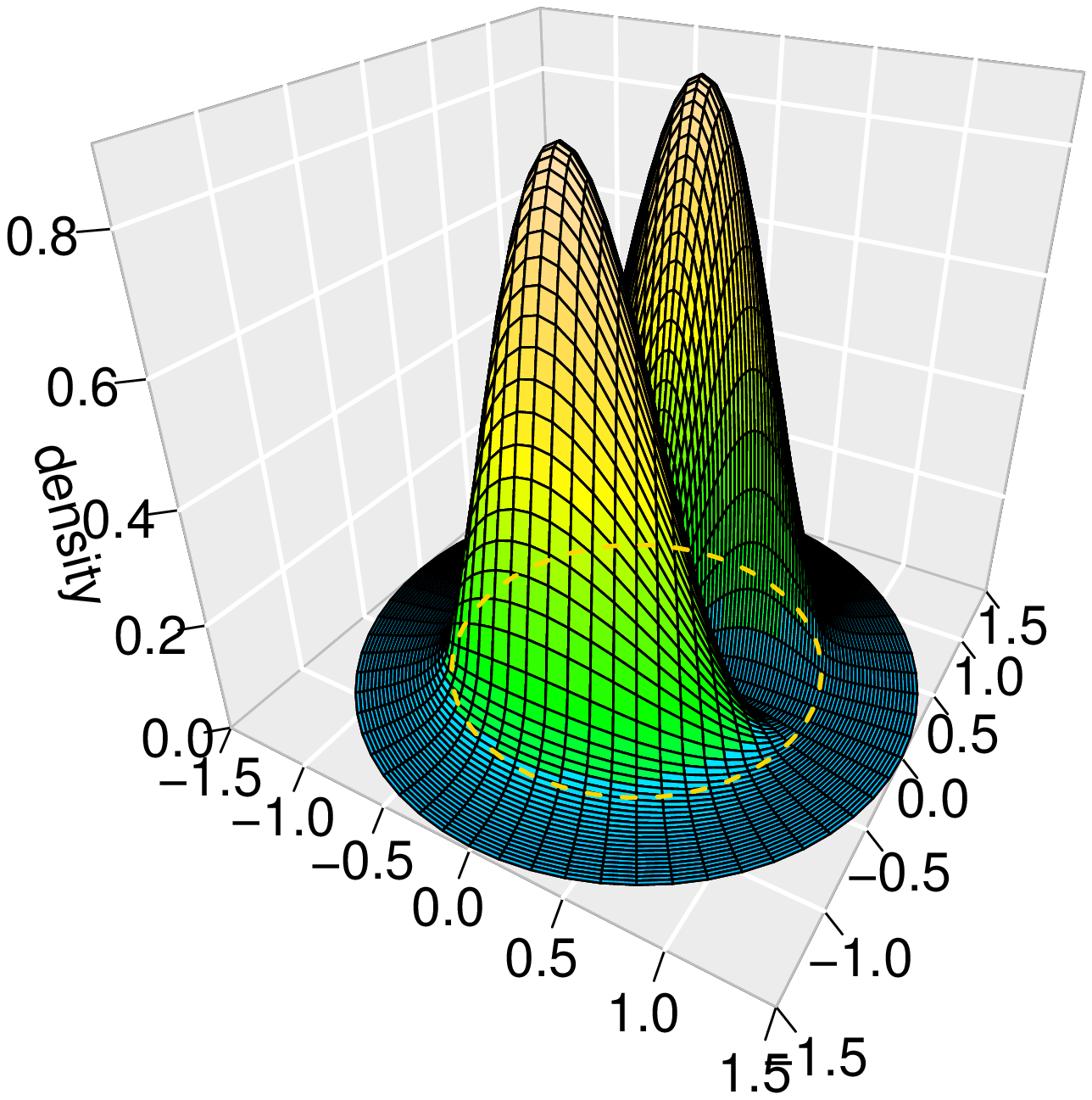}
\includegraphics[width=55mm,height=55mm]{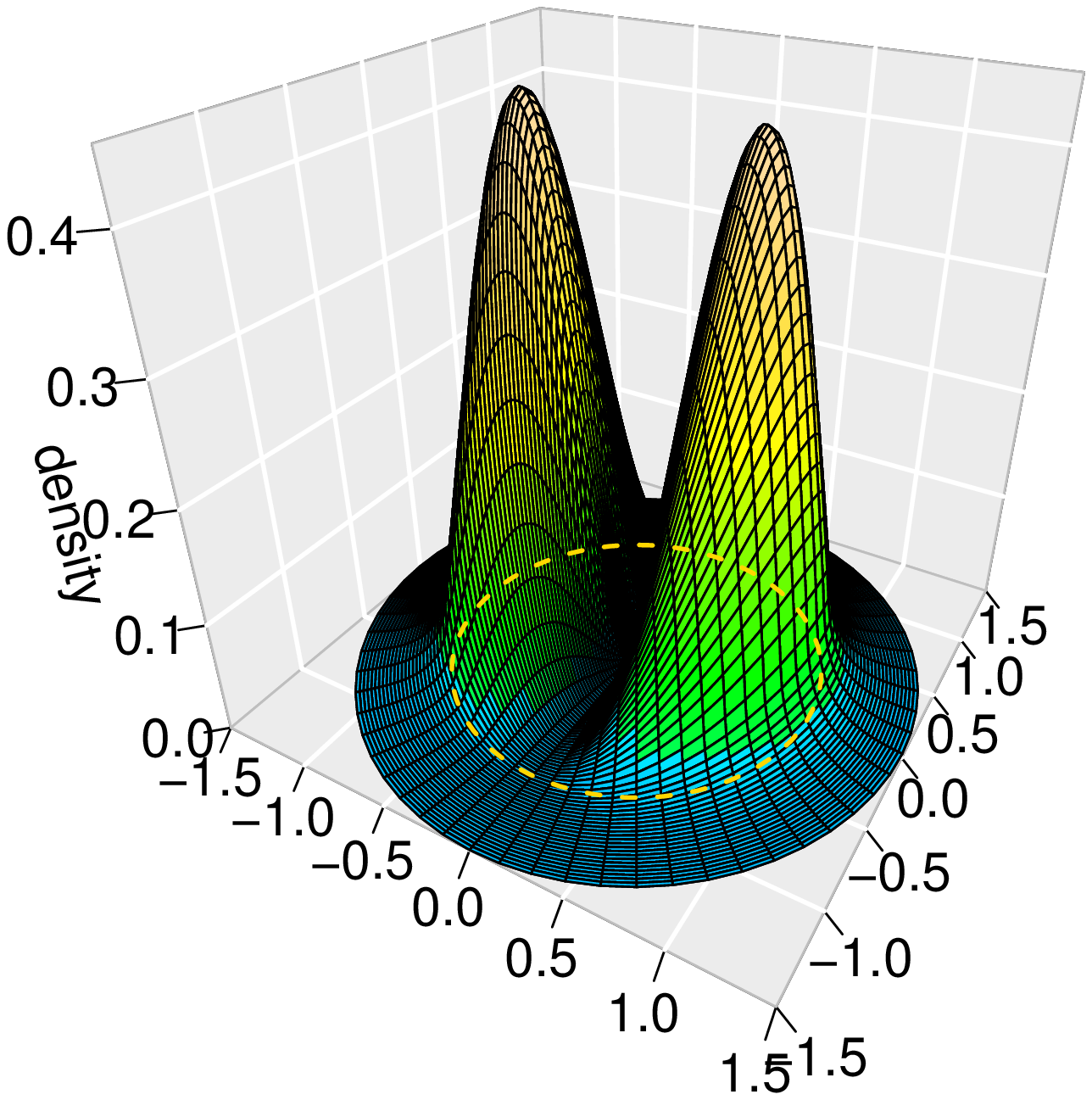}
\caption{$V_0=3.5$:  the radial section of  $\psi_{(1,1,0)}$ and probability density plots for
 $|\psi_{(1,1,0)}(r,\theta,\phi)|^2$,    $|\psi_{(1,1,\pm 1)}(r,\theta,\phi)|^2$.
  We point out that $|Y_{(l,m)}|$  is $\phi $-independent.}
\end{center}
\end{figure}

Quite analogously, we can demonstrate that, if $x_3 f(p)$ is an eigenfunction with the eigenvalue  $E_{(1,1)}^{(V_0=3.5)}=3.46036<3.5$, then
 $x_1 f(p)$  and  $x_2 f(p)$ share with  $x_3 f(p)$ the same $f(p)$,  being  likewise the $E_{(1,1)}^{(V_0=3.5)}$    eigenfunctions.
  The eigenvalue is triply degenerate. Following the observations of Ref. \cite{ZG1} we expect that it is possible express the finite well
   eigenfunctions in terms of  spherical harmonics.  Indeed, \cite{Gr},  we have  $\psi_{(1,1,\pm 1)}(\textbf{x})=(x_1\pm i x_2)f(p)= Y_{1,\pm1} $
   and $\psi _{(1,1,0)}= x_3 f(p)=  Y_{(1,0)} f(p)$.

The eigenvalue problem (26) admits other solutions, that can be retrieved by means of the Strang algorithm.
 We are interested in fairly shallow wells, hence it suffices to mention the next (excited orbital  level)   $l=1$  eigenfunction.
  It exists for $V_0=6.7$, while for  $V_0=6.6$  we have proved the non-existence of eigenfunctions in
  the form  $x_3f(p)$.
 The  $V_0=6.7$  eigenvalue has an approximate value $6.61546$, obtained for  $a=50$ and  $\Delta x=0.001$.
 Its  $a\to \infty $  interpolation (c.f. remark 5) equals  $(6.61546 + 2\cdot(0.0064+0.0032+0.0019+0.0013))=6.64106<6.7$.

\begin{table}[h]
\begin{center}
\begin{tabular}{|c||c|c|}
\hline
a & $E_{(1,1)}$ ($V_0=3.5$) & $E_{(2,1)}$ ($V_0=6.7$) \\
\hline
\hline
50 & 3.43477 & 6.61546  \\
\hline
$\infty$ & 3.46036 & 6.64106  \\
\hline
\end{tabular}
\end{center}
\caption{First two eigenvalues in the $l=1$ series  are admissible once the  threshold value reaches $V_0=6.7$.}
\end{table}

\begin{figure}[h]
\begin{center}
\centering
\includegraphics[width=55mm,height=55mm]{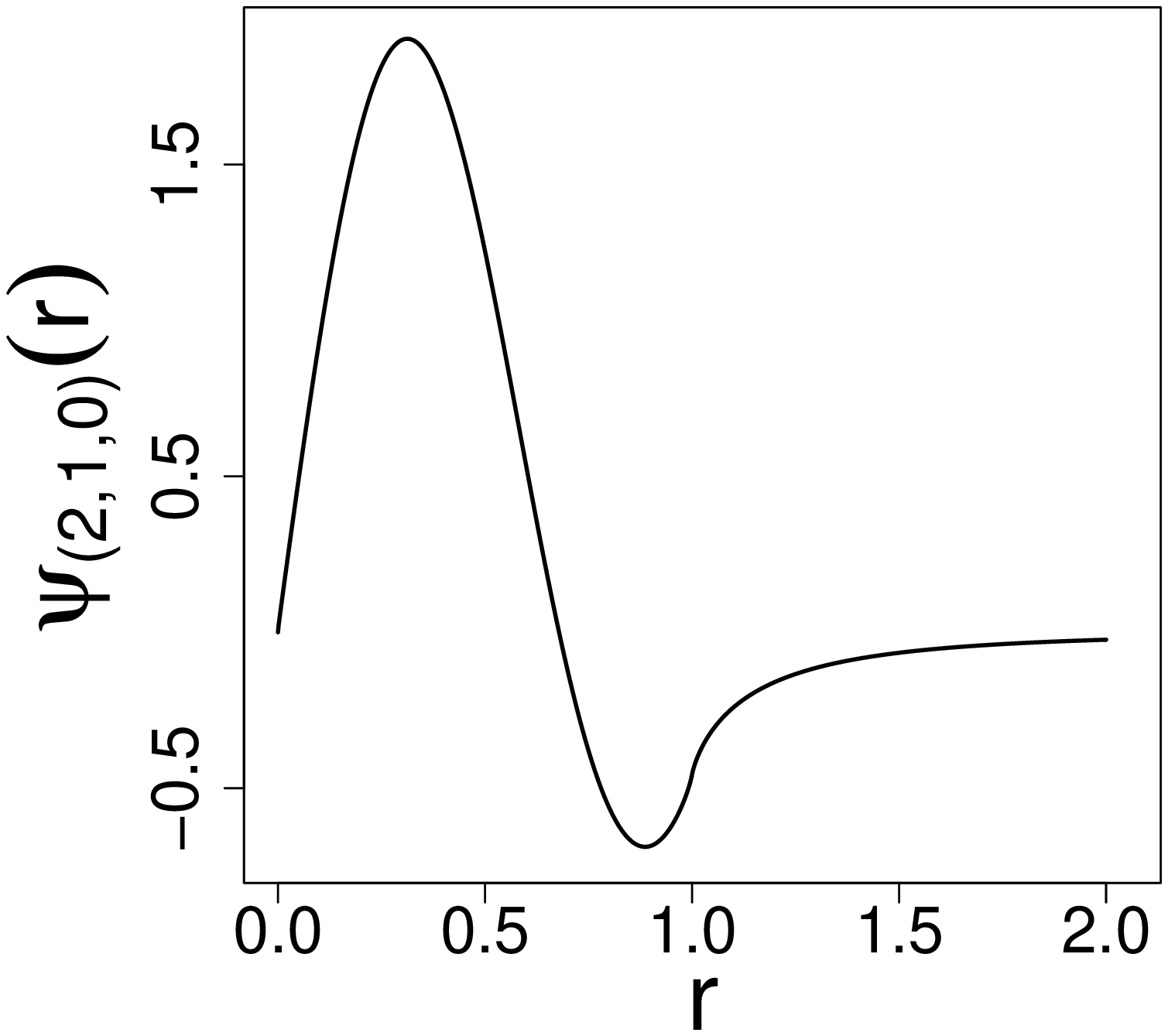}
\includegraphics[width=55mm,height=55mm]{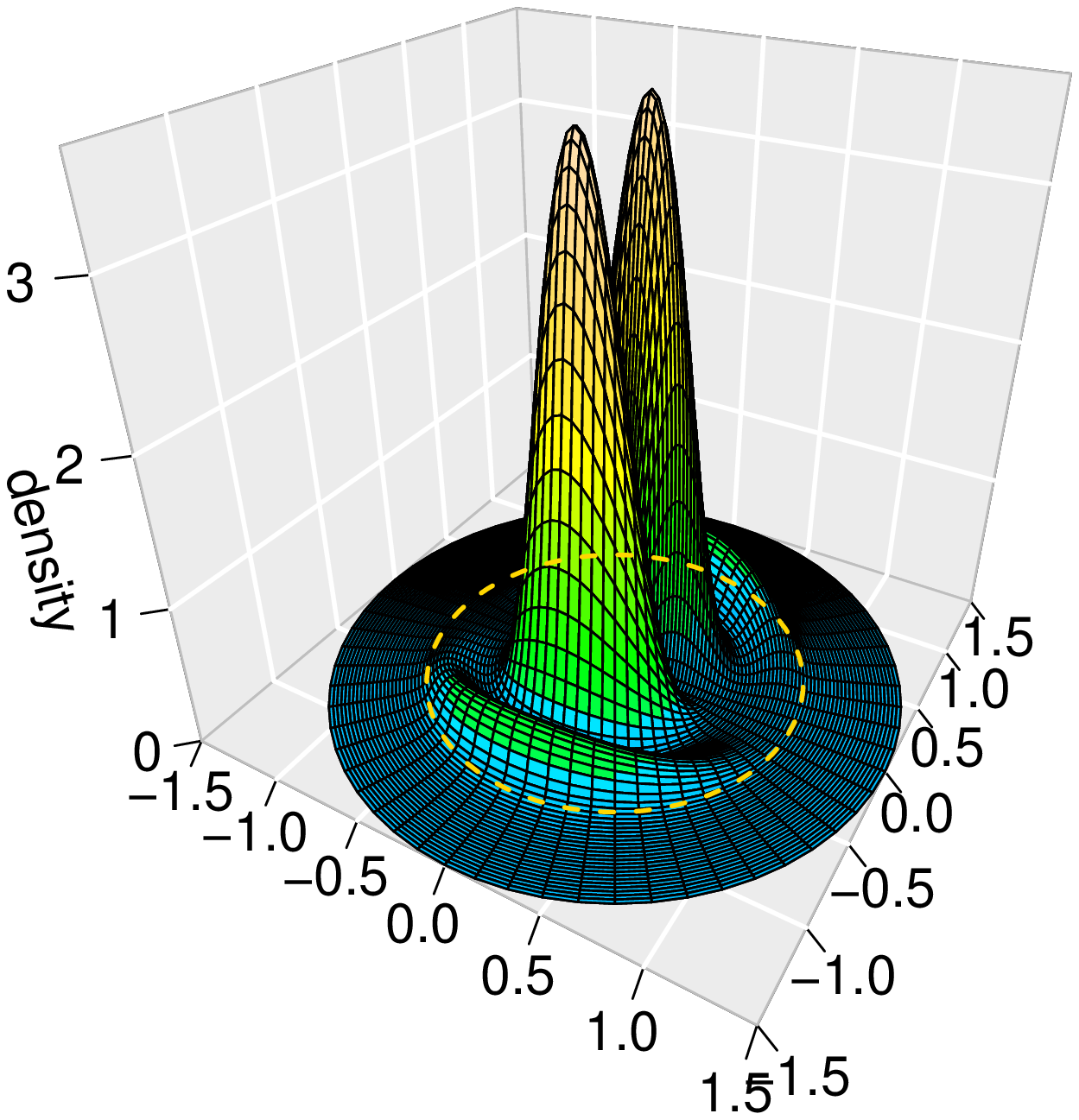}
\includegraphics[width=55mm,height=55mm]{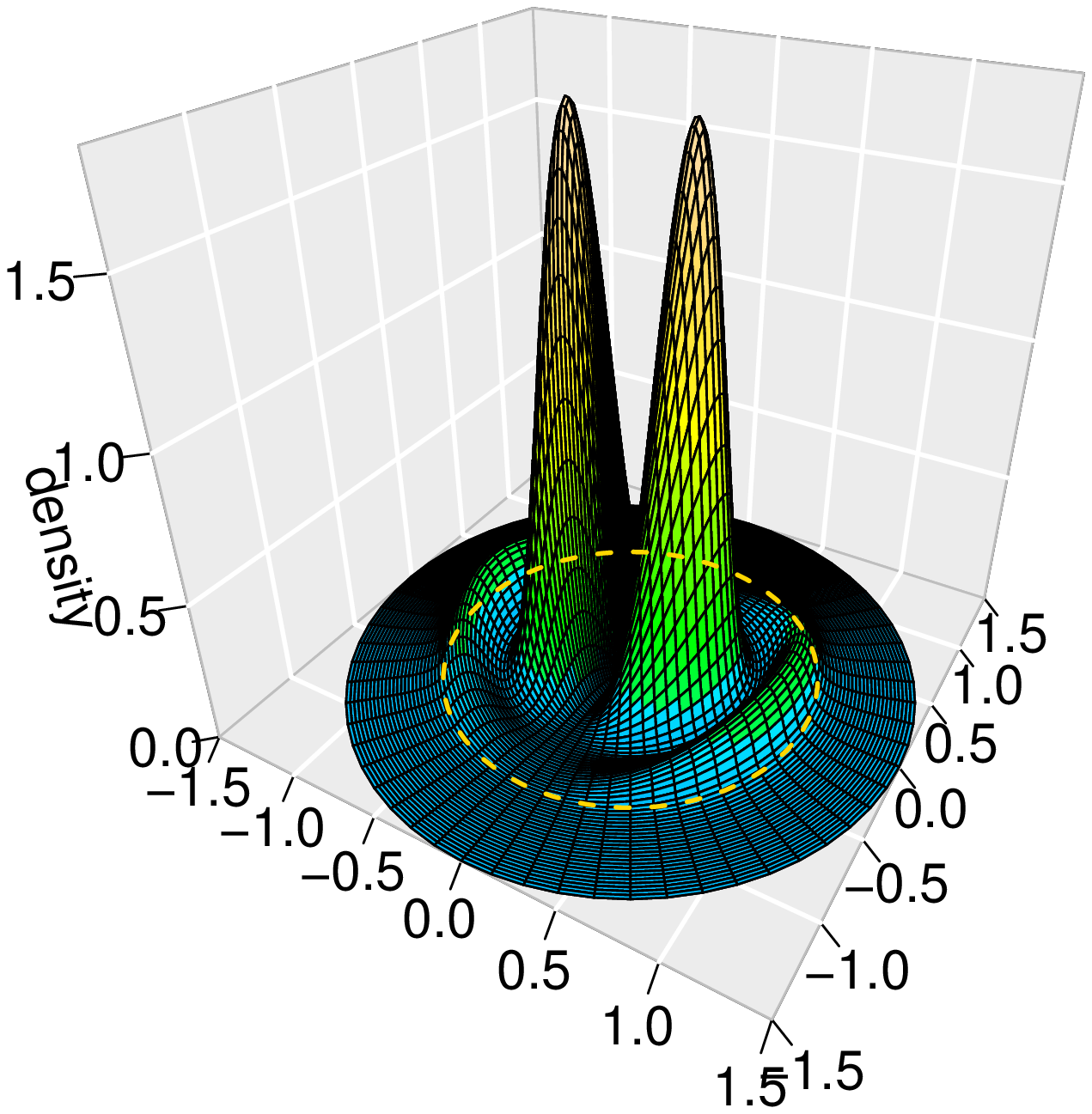}
\caption{The radial section  of  $\psi_{(2,1,0)}$  and polar plots of   probability densities
$|\psi_{(2,1,0)}(r,\theta,\phi)|^2$,   $|\psi_{(2,1,\pm 1)}(r,\theta,\phi)|^2$  for  $V_0=6.7$.}
\end{center}
\end{figure}

\subsection{$l=2$ eigenfunctions}

With the increase of the well depth (or height) above $V_0=3.5$ the $l=2$  orbital eigenfunctions are allowed to appear.
We shall now pass to   $l=2$ proper, while  presuming the  factorization
$\psi_{(k,l,m)}(r,\theta,\phi)=f_k(r)Y_{l,m}(\theta,\phi)$ of eigenfunctions.

We shall demonstrate that  $\psi_{1,2,0}(\textbf{x})=(3x_3^2-p^2)f(p)$, with  $p=\sqrt{\iksy}$, $p\neq 0$   actually is an
orbital solution of Eq. (18).    Integrations are carried out in a fashion similar to that corresponding to the case $l=1$.
 The expression  $A\psi (\bf{p})$ is decomposed into  (remember about the $(p.v.)$ recipe) the difference  $I_1 -I_2$,  where
 \be
\begin{split}
I_1(p)&=\frac{1}{\pi^2}\int\limits_{\mathbb{R}^3}\frac{(3x_3^2-p^2)f(p)d^3u}{((u_1-x_1)^2+(u_2-x_2)^2+(u_3-x_3)^2)^2}=
\frac{(3x_3^2-p^2)f(p)}{\pi^2}\int\limits_{\mathbb{R}^3}\frac{d^3v}{(v_1^2+v_2^2+(v_3-p)^2)^2}\\
&=\frac{(3x_3^2-p^2)f(p)}{\pi^2}\calka \frac{r^2\sin\theta}{(r^2+p^2-2r p\cos\theta)^2}=\frac{(3x_3^2-p^2)f(p)}{\pi p}\int\limits_0^\infty r\left(\frac{1}{(r-p)^2}-\frac{1}{(r+p)^2}\right)dr,
\end{split}
\end{equation}
and
\be
I_2(p)=\frac{1}{\pi^2}\int\limits_{\mathbb{R}^3}\frac{(3u_3^2-u^2)f(u)d^3u}{((u_1-x_1)^2+(u_2-x_2)^2+(u_3-x_3)^2)^2}=
\frac{1}{\pi^2}\int\limits_{\mathbb{R}^3}\frac{[3(s_{11}v_1+s_{12}v_2+s_{13}v_3)^2-v^2]f(v)d^3v}{(v_1^2+v_2^2+(v_3-p)^2)^2}.
\ee
$I_2(p)$ may be given a form of the sum   $I_2=I_{21}+I_{22}+I_{23}+I_{24}$, where
\be
\begin{split}
I_{21}&=\frac{3s_{11}^2}{\pi^2}\int\limits_{\mathbb{R}^3}\frac{v_1^2f(v)d^3v}{(v_1^2+v_2^2+(v_3-p)^2)^2}
=\frac{3s_{11}^2}{\pi^2}\calka\frac{r^2\sin\theta(r\cos\phi\sin\theta)^2f(r)}{(r^2+p^2-2rp\cos\theta)^2}\\
&=\frac{3s_{11}^2}{\pi}\int\limits_0^\infty \left[-\frac{r^2}{p^2}-\frac{r(r^2+p^2)}{4p^3}\left(\ln(r-p)^2-\ln(r+p)^2\right)\right]f(r)dr,
\end{split}
\end{equation}
\be
\begin{split}
I_{22}&=\frac{3s_{12}^2}{\pi^2}\int\limits_{\mathbb{R}^3}\frac{v_2^2f(v)d^3v}{(v_1^2+v_2^2+(v_3-p)^2)^2}
=\frac{3s_{12}^2}{\pi}\int\limits_0^\infty \left[-\frac{r^2}{p^2}-\frac{r(r^2+p^2)}{4p^3}\left(\ln(r-p)^2-\ln(r+p)^2\right)\right]f(r)dr,
\end{split}
\end{equation}
\be
\begin{split}
I_{23}&=\frac{3s_{13}^2}{\pi^2}\int\limits_{\mathbb{R}^3}\frac{v_3^2f(v)d^3v}{(v_1^2+v_2^2+(v_3-p)^2)^2}\\
&=\frac{3s_{13}^2}{2\pi p^3}\int\limits_0^\infty \left[r(r^4+p^4)\left(\frac{1}{(r-p)^2}-\frac{1}{(r+p)^2}\right)+
r(r^2+p^2)\left(\ln(r-p)^2-\ln(r+p)^2\right)\right]f(r)dr,
\end{split}
\end{equation}
\be
\begin{split}
I_{24}&=-\frac{1}{\pi^2}\int\limits_{\mathbb{R}^3}\frac{v^2f(v)d^3v}{(v_1^2+v_2^2+(v_3-p)^2)^2}
=-\frac{1}{\pi p}\int\limits_0^\infty r^3f(r)\left(\frac{1}{(r-p)^2}-\frac{1}{(r+p)^2}\right)dr.
\end{split}
\end{equation}
We deal with singular integrals hence the $I = (p.v.)(I_1 - I_2)$  recipe must be kept in mind.
Collecting all terms together we arrive at   $I(p)=(3x_3^2-p^2)(Af)(p)$,
where $(Af)(p)$  of Eq. (26)  has the  $(p.v.)$ form
\be
(Af)(p) =
\ee
$$
 \frac{1}{4\pi}\int\limits_0^\infty \left[\frac{r}{p}\left (4f(p)-f(r)\frac{3(r^4+p^4)-2r^2p^2}{p^4}\right)
 \left(\frac{1}{(r-p)^2} - \frac{1}{(r+p)^2}\right)  - 3f(r)\frac{r(r^2+p^2)}{p^5}\left(\ln(r-p)^2-\ln(r+p)^2\right)\right]dr.
$$

Like in case of $l=1$ where the  factor $x_3$ has been spurious, we identify  $(3x_3^2-p^2)\neq 0$ as
the spurious factor,  thus reducing the eigenvalue problem to the form (26).
  The Strang algorithm is employed again, with the assumption about the
 normalization of $\psi $
 \be
C^2\calka r^2\sin\theta(3r^2\cos^2\theta-r^2)^2f^2(r)=\frac{16\pi C^2}{5}\int\limits_0^\infty r^6f(r)dr=1,
\ee
and  the scalar product   $<f_1|f_2>$   directly inferred from $<\psi _1|\psi _2>$, under an assumption that $\psi = (3x_3^2-p^2)f(p)$:
\be
\calka r^2\sin\theta(3r^2\cos^2\theta-r^2)^2f_1(r)f_2(r)=\frac{16\pi}{5}\int\limits_0^\infty r^6f_1(r)f_2(r)dr.
\ee

We have no clues about the threshold $V_0$ value above which the first $l=2$ eigenstate does appear. Our reasoning was somewhat empirical
(e.g. via numerical guesses  and  tests).  With $a=50$  we have found the for   $V_0=4.8$ the orbital $l=2$  eigenstate exists, while
for $V_0=4.7$ there is none.

\begin{figure}[h]
\begin{center}
\centering
\includegraphics[width=80mm,height=80mm]{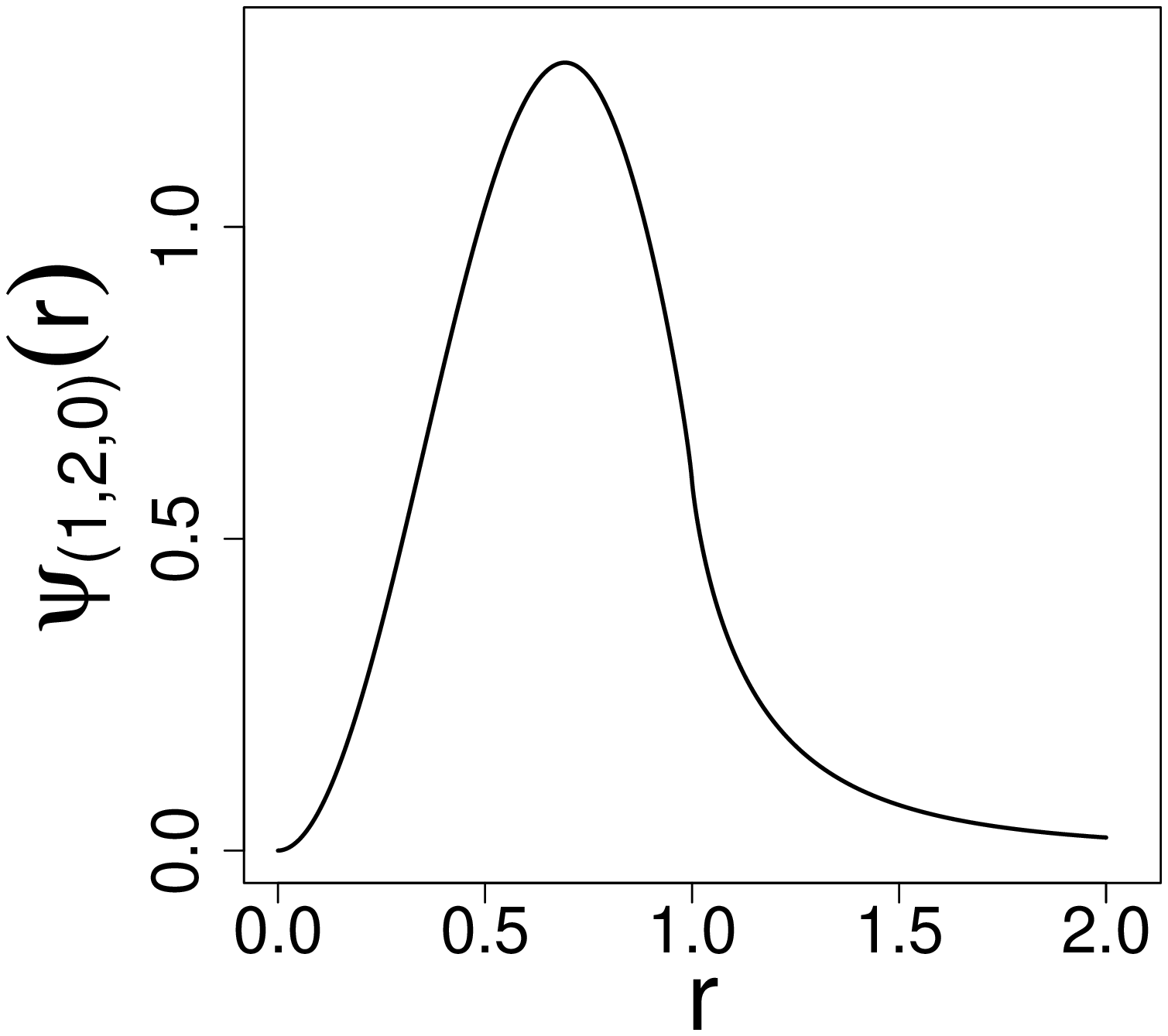}
\includegraphics[width=80mm,height=80mm]{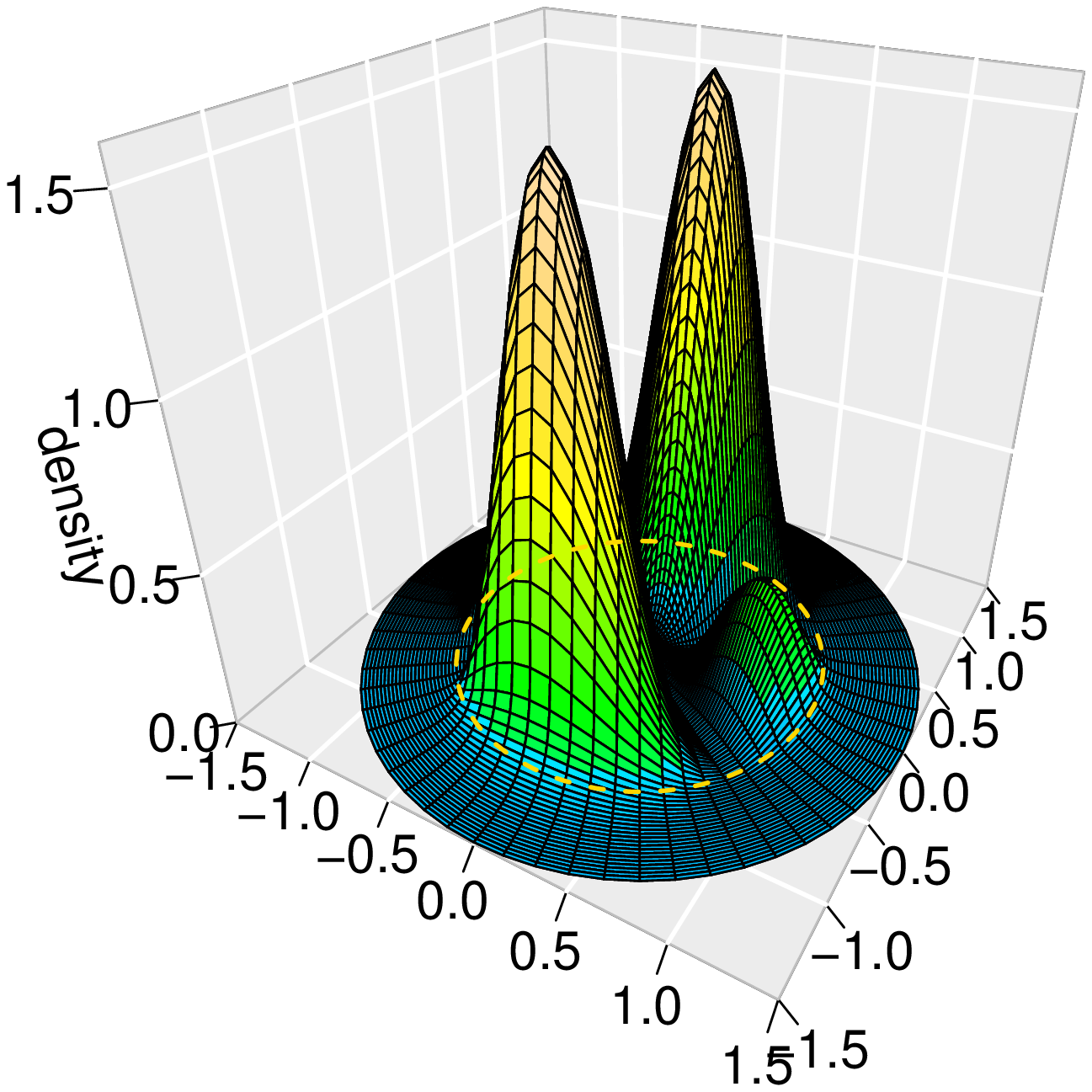}
\caption{$V_0=4.8$: the radial section of $\psi_{(1,2,0)}$ and polar plots for  the  probability density
 $|\psi_{(1,2,0)}(r,\theta,\phi)|^2$.}
\end{center}
\end{figure}
\begin{figure}[h]
\begin{center}
\centering
\includegraphics[width=80mm,height=80mm]{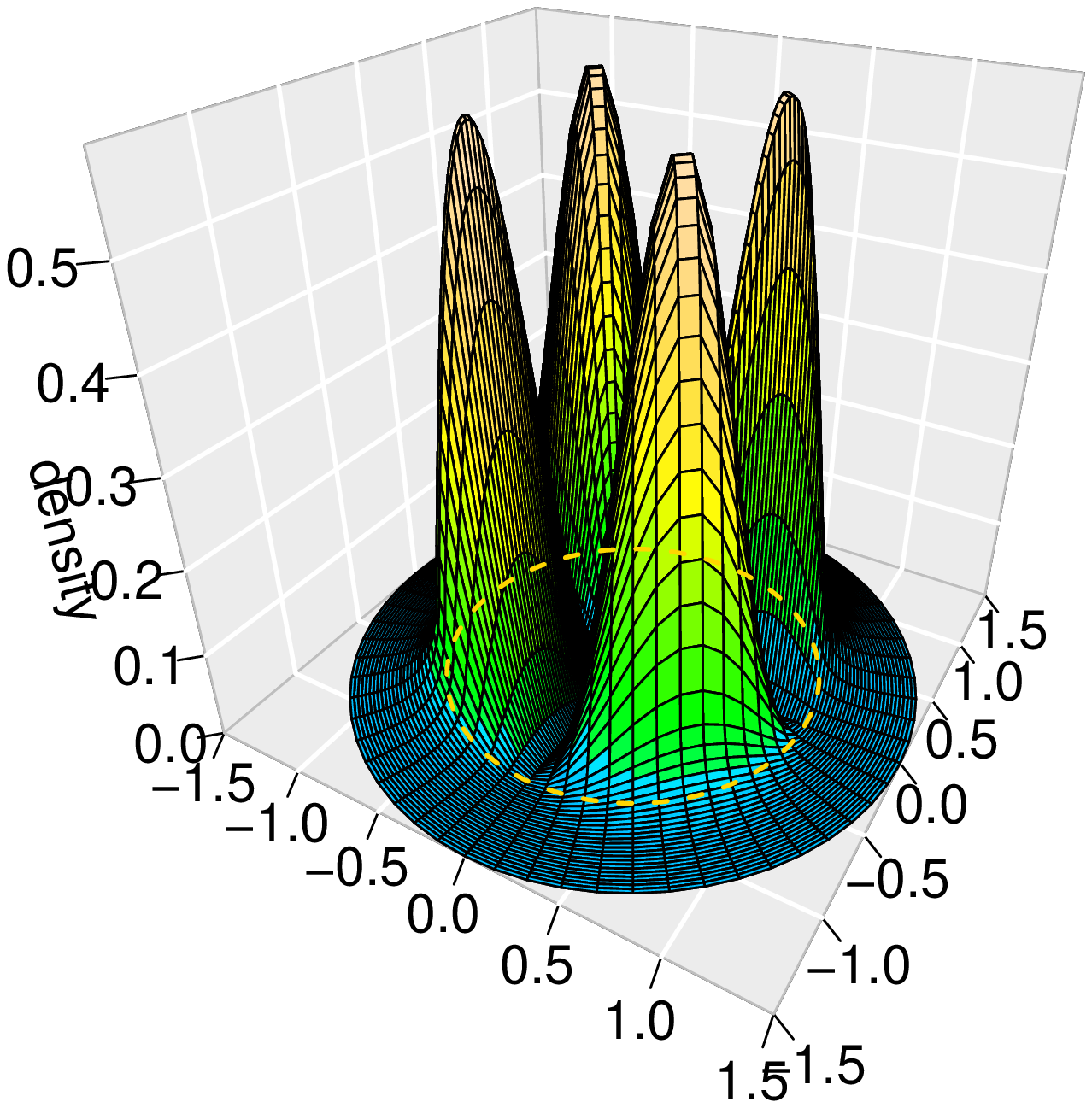}
\includegraphics[width=80mm,height=80mm]{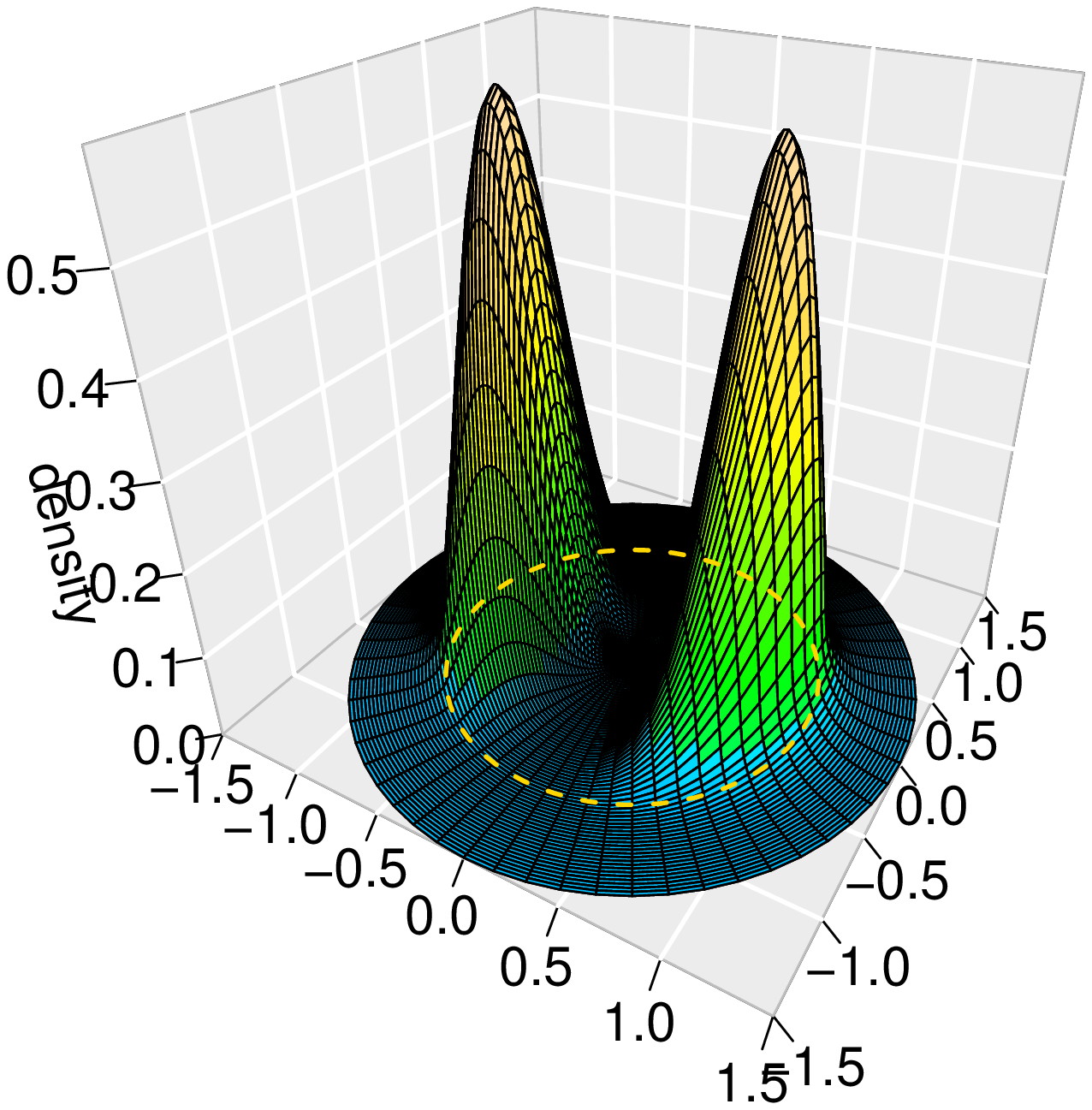}
\caption{Polar plots of probability densities  $|\psi_{(1,2,\pm 1)}(r,\phi,\theta)|^2$ and
 $|\psi_{(1,2,\pm 2)}(r,\phi,\theta)|^2$ for  $V_0=4.8$.}
\end{center}
\end{figure}

\begin{table}[h]
\begin{center}
\begin{tabular}{|c||c|}
\hline
a & E ($V_0=4.8$) \\
\hline
\hline
50 & 4.74184\\
\hline
100 & 4.75463\\
\hline
200 & 4.76102\\
\hline
500 & 4.76486\\
\hline
$\infty$ & 4.76746\\
\hline
\end{tabular}
\end{center}
\caption{$V_0=4.8$: the $a$-dependence of the radial eigenvalue $E_{(1,2)}$.}
\end{table}
We can evaluate the energy intervals (gaps) between consecutive $a$-dependent  eigenvalues.
They read  $0.01279, 0.00639, 0.00384$ and are roughly (up to the last decimal digit) doubled  $1D$ gaps of Remark 5.
Accordingly, we anticipate the  $a\to\infty$  eigenvalue by taking $E(a=500)$  and renormalizing it (additively) by
 $2\cdot 0.0013$. That implies  $E_{(1,2)}^{V_0=4.8}=4.76746$.\\

One can verify that functions of the form
\be
(x_1^2-x_2^2)f(p),\quad x_1x_2f(p),\quad x_1x_3f(p),\quad x_2x_3f(p),
\ee
are  eigenfunctions corresponding to the  common  with  $\psi_{1,2,0}(\textbf{x})=(3x_3^2-p^2)f(p)$
eigenvalue   $E_{(1,2)}$ of Table V.
An equivalent eigenfunction set displays a manifest dependence on spherical  harmonics, in consistency with the infinite well
 observations of Ref. \cite{ZG1}:
\be
\begin{split}
\psi_{(1,2,0)}(p)&=C(3x_3^2-p^2)f(p)=\tilde{C}p^2Y_2^0f(p),\\
\psi_{(1,2,\pm 1)}(p)&=C x_3(x_1\pm i x_2)f(p)=\tilde{C}p^2Y_2^{\pm 1}f(p),\\
\psi_{(1,2,\pm 2)}(p)&=C(x_1\pm i x_2)^2f(p)=\tilde{C}p^2Y_2^{\pm 2}f(p).
\end{split}
\ee

By collecting together the obtained spectral data, we can tabulate the threshold  well  height (depth)
 values $V_0$  so that the maximal number
 of bound states can be clearly  identified.  To this end we provide  a cumulative table (Table VI)
  of  computed  eigenvalues  for $2.1\leq V_0 \leq 8.3 $, comprising  {\it all}  $0\leq l\leq 2$ entries.
We  keep intact  the  presumed $0.1$ inaccuracy  with which a given  existence threshold is located, i.e.
given the depicted  $V_0$  value, the actual threshold  is  located within the  interval $(V_0 - 0.1, V_0]$.

\begin{table}[h]
\begin{center}
\begin{tabular}{|c|c|c|c|c|c|c|c|c|c|}
\hline
$V_0$ & 2.1 & 3.5 & 4.8 & 5.2 & 6.7 & 8.1 & 8.3 & $\ldots$ & $\infty$ \\
\hline
\hline
$l=0$  & 2.03882 &  2.27399 & 2.37186 & 2.39313 & 2.45288 & 2.49118 & 2.49575 &\ldots & $2.754 769 = E_{(1,0)}$\\
 & - &-  &-  & 5.14626  & 5.31977 & 5.40211 & 5.41186 & \ldots & $5.892 214 = E_{(2,0)}$ \\
 & - &-  &-  & - & - & - & 8.28013 & \ldots & $9.033 009 = E_{(3,0)}$  \\
\hline
$l=1$ & - & 3.46036 & 3.60459 & 3.63433 & 3.71631 & 3.76785 & 3.77396 & \ldots  & $4.121 332 = E_{(1,1)}$\\
 & - & - & - & - & 6.64106 & 6.75640 & 6.76877  & \ldots  & $7.342 181 = E_{(2,1)}$\\
\hline
$l=2$ & - & - & 4.76746 & 4.80535 &  4.90712 & 4.96999 & 4.97740 & \ldots & $5.400 079 = E_{(1,2)}$\\
 & - & - & - & - & - & 8.04169 & 8.05689  & \ldots & $8.718 436 = E_{(2,2)}$\\
\hline
\end{tabular}
\end{center}
\caption{Computed eigenvalues  (interpolated in accordance with the $a\to\infty$ recipe  of Remark 5)  for various  $V_0$ choices.
The last column, reproducing the corresponding  infinite spherical well eigenvalues $E_{k,l)}$,
 has been borrowed  from Table III  in Ref. \cite{ZG1}.}
\end{table}

In Table VI, we can read out a maximal number of admitted $l\leq 2$ eigenvalues,
together with an order according to which the
consecutive eigenvalues are allowed to appear  with the growth of  $V_0$.

We indicate that $E_{(1,3)}= 6.630371$, \cite{ZG1},  hence one  can expect
 an  emergence of the first $l=3$ eigenvalue  in the finite well  at $V_0$ around $5.7 -  5.9$.  In view of
 $E_{(2,3)} = 10.045716$, in the finite well
the second $l=3$ eigenvalue could possibly appear for $V_0$  about $9.2 -9.3$.
Accordingly, Table VI  up to $V_0=5.2$ depicts a maximal number of admitted eigenvalues,
 while for $6.7 \leq V_0 \leq 8.3$ only a single (actually  the  first) $l=3$ explicit eigenvalue
 is missing in the Table.

\section{Outlook}

We recall that  the basic goal of  Ref. \cite{gar0} was to  set on solid grounds the quantization programme
which   completely avoids any  reference to a classical mechanics of massive particles,  traditionally
 viewed as a conceptual support for the  choice of the Hamiltonian operator within the standard  Schr\"{o}dinger wave mechanics.
We have indicated   there  that a commonly adopted  form of the Hamiltonian
(minus Laplacian plus a perturbing potential)  an exception rather than a universally valid feature
of an admissible quantum theory, for which the choice  of  $- (\Delta )^{\alpha /2}$, $\alpha \in (0,2)$,
 instead of  $-\Delta $  does not at all exhaust  an infinity of other candidate operators
 (c.f. the L\'{e}vy-Khintchine formula,  \cite{gar0}) .

In the present paper our focus was on   spectral properties  of the bound states in the  ultrarelativistic case.
That amounts to  a  concrete  choice of  $\alpha =1$, resulting in the  Cauchy operator
  $\sqrt{-\Delta}$. It  is  perhaps the only  directly physics-motivated example, which can be
    singled out  from  the  one-parameter   family of L\'{e}vy stable operators
     $(-\Delta)^{\alpha /2}$.    Each of these operators gives rise  to a family of related
     Schr\"{o}dinger-type   spectral problems, see e.g.  \cite{carmona}.

   The absence of any {\it natural} mass parameter is a conspicuous feature of
   L\'{e}vy-Schr\"{o}dinger spectral problems and   their   ultrarelativistic (Cauchy)  version
    in particular.

There are  few only   spectral solutions  that have been  obtained in the ultrarelativistic  regime,
with the main activity arena being  $1D$.  In particular, the $1D$ Cauchy oscillator problems
 has been  analytically  solved in Refs. \cite{gar,lorinczi2}.
  Its anharmonic version  has been  addressed in ref. \cite{lorinczi3}.
 The $3D$   Cauchy oscillator has been addressed     in \cite{remb} in the $l=0$ sector, hence
  with no  orbitally nontrivial outcomes.

  We have contributed to an active research on Cauchy operators with exterior Dirichlet boundary data  in $1D$
   (infnite well problem), extending that  analysis to the finite well spectral problem,
    \cite{ZG2,ZG3}, see also  \cite{duo,dyda0} and \cite{GZ}.

In the previous paper \cite{ZG1}, we have addressed the general  $3D$ spectral problem for the
 ultrarelativistic spherical infinite well, see e.g. \cite{dyda0,dyda}  for  related considerations.
 Presently our focus was on the more physically appealing  case of the  finite spherical well.

We have discussed, in part with the aid of numerical methods,  the existence issue for the ground state.
Next, approximate threshold values for the emergence of higher excited  states were established,
both in the radial and nontrivially orbital sectors.
The corresponding eigenfunction shapes were established as well, while accounting for the degeneracy
 of the spectrum.
The eigenvalues obtained under the shallow well premises were  collected in the  Table VI,  and set in correspondence
 (albeit somewhat  distant)  with those for the infinite $3D$ well.

We end up with a potentially interesting research hint of Ref. \cite{ZG1}:
 "quite  an ambitious research   goal  could be an analysis a  spatially
  random distribution ("gas") of finite ultrarelativistic  spherical wells,   embedded in a  spatially extended finite
  energy background".
   This  could be further source of inspirations in   attempts  towards  understanding  how (possibly
    on  large  spatial scales) the   energy can  be    stored or accumulated  in the form of  bound
    states of Schr\"{o}dinger - type  quantum systems, that are devoid of any mass. \\

   {\bf Ackonwledgement:} We would like to express our  gratitude to Professor Jozsef L\H{o}rinczi for illuminating
   exchanges   on various aspects of  finite well spectral problems for fractional Laplacians, c.f. \cite{thanks} in the text.

\end{document}